\documentclass{aa}  
\usepackage{graphics,graphicx,lipsum}
\usepackage{wrapfig}
\usepackage{caption}
\usepackage{subcaption}
\usepackage{amsmath}
\usepackage{amssymb}
\usepackage{color, xcolor}
\usepackage[utf8]{inputenc}
\usepackage{natbib}
\usepackage{enumitem}
\usepackage{xparse}
\usepackage{float}
\usepackage{dblfloatfix}
\restylefloat{table}
\usepackage{placeins}
\usepackage{diagbox}
\usepackage{tablefootnote}
\usepackage[flushleft]{threeparttable}
\usepackage{txfonts}
\usepackage{natbib}
\usepackage{scrextend}
\usepackage{hyperref}
\usepackage{cleveref}
\usepackage{tabularx}
\usepackage{booktabs}
\usepackage{array}
\usepackage{rotating}

\def\Hii{H\,{\sc ii}}

\def\cmss{cm\,s$^{-2}$}

\def\msun{M$_{\odot}$}
\def\mstar{M$_\star$~}

\def\vsini{$v \sin i$}

\def\Msun{M$_\odot$}

\def\micron{$\mu$m}
\def\thirteenco{{$^{13}$CO}}
\def\first{$1^{\rm st}$}
\def\second{$2^{\rm nd}$}

\begin{document}

   \title{Massive pre-main-sequence stars in M17}

   \subtitle{\first\,and \second\,overtone CO bandhead emission and the thermal infrared}

   \author{\mbox{J. Poorta}\inst{1}
         \and
         \mbox{M.C. Ram\'irez-Tannus}\inst{3} 
         \and
         \mbox{A. de Koter}\inst{1,2}
         \and
         \mbox{F. Backs}\inst{1}
         \and
         \mbox{A. Derkink}\inst{1}
         \and
         \mbox{A. Bik}\inst{4}
         \and
         \mbox{L. Kaper}\inst{1}
    }

   \institute{
        Anton Pannekoek Institute for Astronomy, University of Amsterdam,
              Science Park 904, 1098 XH Amsterdam, The Netherlands\\
              \email{j.poorta@uva.nl}
        \and
        Institute of Astrophysics, Universiteit Leuven, Celestijnenlaan 200 D, 3001 Leuven, Belgium
        \and
        Max Planck Institute for Astronomy, Königstuhl 17, 
            D-69117 Heidelberg, Germany
        \and 
        Department of Astronomy, Oskar Klein Centre, Stockholm University, AlbaNova University Centre, 106 91, Stockholm, Sweden
    }

    \date{Received 9 December 2022 / Accepted 19 April 2023}

 
  \abstract
{Recently much progress has been made in probing the embedded stages of massive star formation, pointing to formation scenarios that are reminiscent of a scaled up version of low-mass star formation. However, the latest stages of massive star formation have rarely been observed, as young massive stars are assumed to reveal their photospheres only when they are fully formed.}
{Using \first\, and \second\, overtone CO bandhead emission and near- to mid-infrared photometry we aim to characterize the remnant formation disks around 5 unique pre-main-sequence (PMS) stars with masses $6-12$ \msun, that have constrained stellar parameters thanks to their detectable photospheres. We seek to understand this emission and the disks it originates from in the context of the evolutionary stage of the studied sources.}
{We use an analytic LTE disk model to fit the CO bandhead and the dust emission, assumed to originate in different disk regions. For the first time we modeled the \second\, overtone emission, which helps us to put tighter constraints on the density of the CO gas. Furthermore, we fit continuum normalized bandheads, using models for stellar and dust continuum, and show the importance of this in constraining the emission region. We also include \thirteenco\, in our models as an additional probe of the young nature of the studied objects.}
{We find that the CO emission originates in a narrow region close to the star ($< 1$ AU) and under very similar disk conditions (temperatures and densities) for the different objects. This is consistent with previous modeling of this emission in a diverse range of young stellar objects and identifies CO emission as an indicator of the presence of a gaseous inner disk reaching close to the stellar surface. 
From constraining the location of the inner edge of the dust emission, we find that all but one of the objects have undisrupted inner dust disks.} 
{We discuss these results in the context of the positions of these PMS stars in the Hertzsprung-Russel diagram and the CO emission's association with early age and high accretion rates in (massive) young stellar objects. We conclude that, considering their mass range and for the fact that their photospheres are detected, the M17 PMS stars are observed in a relatively early formation stage. They are therefore excellent candidates for longer wavelength studies to further constrain the end stages of massive star formation.}

  \maketitle

\section{Introduction} \label{sec:introduction}
The impact massive stars (\mstar $ \geq  8$ \msun) have on their host galaxies is disproportionate to their numbers: they provide strong mechanical and radiative feedback, and deposit new elements in their surroundings through powerful stellar winds and supernova ejecta which constitute the building blocks of planets and life. Understanding the way in which these stars form is therefore important, yet challenging. The formation events are rare, take place deep inside a dust obscured natal environment and unfold on such short timescales that the stars arrive on the main sequence prior to the dispersal of their natal cloud. Notwithstanding these complicating factors, both observational \citep[e.g.][]{johnston2015,ilee2016,cesaroni2017,beuther2018a} and theoretical \citep[e.g.][]{krumholz2009, kuiper2010,rosen2016,meyer2018} evidence is mounting that disk mediated accretion is crucial to the process. 

Recent studies that reveal disks, large-scale disk-like structures and/or outflows, probe the embedded stages of formation \citep[e.g.][]{frost2019,frost2021a,maud2018,pomohaci2017}. However, much remains unclear about how massive stars arrive on the main sequence, what their stellar properties are when they do, and how and when their accretion is halted. One of the ways to study the latest stages of formation is to observe newly formed massive stars in their natal environment. By making a census of the young populations in massive star forming regions and determining their stellar and circumstellar (e.g. multiplicity, disks) characteristics, one can constrain possible formation scenarios and provide robust initial conditions for population synthesis studies \citep[e.g.][]{bik2006}. Through one such study in the very young ($\leq 1$ Myr) Galactic cluster M17, \citet[hereafter RT17]{ramirez-tannus2017} made the unique discovery of 6 PMS stars in the mass range $\sim 6-12$ \msun, that have observable photospheres and at the same time show spectroscopic and photometric features typical for more embedded massive young stellar objects (MYSOs).

RT17 obtained optical to near-infrared (NIR) spectra with VLT/X-shooter of 12 young OB stars with masses ranging between $\sim 6-25$ \msun. For most of these objects the photospheric part of the spectrum was of sufficient quality to allow quantitative spectroscopic modeling to determine stellar properties. Combining this with available photometry they confirmed the pre-main-sequence (PMS) nature of 6 objects ($\sim 6-12$ \msun) in the sample, based on their position in the Hertzsprung-Russell diagram (HRD; see \Cref{sec:kindofdisk} and \Cref{fig:hrd}). Two more objects (B289 and B215) were found to be close to, but not on, the zero age main sequence, displaying only mid-infrared (MIR) excess and lacking the other disk signatures listed below. These objects were characterized by RT17 as young stars whose PMS nature could not be solidly confirmed. The objects that \emph{were} confirmed to be PMS stars possess features which point to the presence of circumstellar disks: (1) double peaked atomic emission lines, most notably H\,{\sc i}, Ca\,{\sc ii} and O\,{\sc i}; (2) CO-bandhead emission; and (3) NIR to MIR excess. Of these, especially CO bandhead emission has been associated with high accretion rates and the earlier stages of formation in young stellar objects (YSOs; see \Cref{sec:kindofdisk} and \Cref{tab:CO_detection_rates} for references).   

The ro-vibrational transitions of the CO molecule that result in the CO bandhead emission, have recurrently been used to study the innermost regions of circumstellar disks in YSOs of all masses. Especially for the higher mass YSOs, where these regions can rarely be resolved, they constitute one of the few commonly used diagnostics. A few studies focus on a single object \citep[e.g.][]{gravitycollaboration2020, fedriani2020}, revealing the spatial distribution of the emission with interferometry or integral field units. Other studies have gathered samples of intermediate to high mass YSOs to study the disk properties in a more statistical manner \citep{bik2004,wheelwright2010,ilee2013,ilee2014}. In all these cases the emission has been modeled either as a flat disk or a ring of material in Keplerian rotation, with the general result that the emission likely originates from hot ($2000-5000$ K) and dense ($N_{\text{CO}} \sim 10^{20} - 10^{22}~\text{cm}^{-2}$) disk regions inside of the dust sublimation radius. 

This paper's focus is to constrain the properties of the circumstellar material for the five objects among the RT17 sample of PMS stars in M17 that show CO bandhead emission. As these objects also reveal NIR to MIR excess, we include this thermal emission in our analysis. Based on these features we aim to determine properties of the inner gaseous disk ($< 1$ AU) where the CO-emission likely originates, and the somewhat further regions ($\geq 3-5$ AU) where the bulk of the thermal NIR and MIR dust emission comes from. Our sample consists of relatively high mass ($6-12$ \msun) PMS objects, \emph{that additionally have well constrained stellar properties}. This allows us to address the question: can the disk properties be linked to stellar properties, such as mass and evolutionary stage? 

Though we take a similar modeling approach to \citet{bik2004}, \citet{wheelwright2010}, \citet{ilee2013} and \citet{ilee2014}, we refine it in several ways (see \cref{sec:model_exploration} for details). First, for the first time, we take into account the observed $2^{\rm nd}$ overtone emission ($ \Delta v = 3 $, from $ 1.54$ \micron), as opposed to only the $1^{\rm st}$ overtone ($\Delta v = 2 $, from $2.29$ \micron). All the included (\first\,and \second\,overtone) bandheads are fit together, rather than individually. Second, by studying the dust emission with the available photometry we obtain robust continuum estimates, which we use to normalize the modeled bandheads. Third, we take into account $^{13}$CO, which increases the quality of our fits and allows us to probe the isotopologue ratio $^{13}$CO/$^{12}$CO as a further test of the young nature of our sample.

The paper is organized as follows. In \Cref{sec:data} we present the data and the previous analysis performed by RT17. The disk model and the treatment of gas and dust are explained in \Cref{sec:methods}. In this section we also explore the sensitivity of the predicted line spectrum to model parameters. In \Cref{sec:results} we describe our fitting approach and the resulting disk parameters. \Cref{sec:discussion} places these results in the context of previous literature and of the evolutionary state of the M17 PMS stars. \Cref{sec:conclusions} contains a brief summary and our main findings.        

\section{Data and previous analysis} \label{sec:data}

\begin{table*}[ht!]
\footnotesize
\centering
\caption{Stellar and extinction properties derived from quantitative spectroscopy and optical ($\lambda \lessapprox 1 \mu$) SED fitting.}       
\begin{minipage}{0.7\hsize}
\centering
\renewcommand{\arraystretch}{1.4}
\setlength{\tabcolsep}{3pt}
\begin{tabular}{lcccccccc}
\hline
\hline
Name & Sp. Type & $T_{\rm eff}$	  &  $\log g$               & $A_V   $  & $\log L/L_{\odot}$ & $R_{\star}$ & $M$\footnote{~ZAMS mass of best fit PMS track.}  & $\rm Age_{\rm HRD}$ \\ 
           &   &  K              &  $\rm cm$\,$\rm s^{-2}$ &            &                   &   $R_{\odot}$     &  $M_{\odot}$   &  Myr \\ 
\hline                                                                                                                                                             
B163$^{b}$ &  kA5  &    8200                  &  $-$                          & $13.2^{\uparrow}_{\downarrow}$&$2.95^{\uparrow}_{\downarrow} $  &  $ 10.1^{\uparrow}_{\downarrow}$ &  6   & 0.14 \\ 
B243 &  B8 V &   $13500^{+1350}_{-1250}$ &  $4.34^{\uparrow}_{-0.3} $   & $8.5^{\uparrow}_{-1.0} $    &$3.21^{+0.07}_{-0.06} $            &$7.5^{+1.0}_{-0.8}$               & 6 & 0.20  \\ 
B268 &  B9-A0 &  $12250^{+850}_{-1000}$  &  $3.99^{\uparrow}_{-0.38}$   & $8.1^{\uparrow}_{-1.0}    $ &$3.24^{+0.04}_{-0.05} $            &$8.8^{+1.2}_{-0.8}$               & 6 & 0.20   \\ 
B275 &  B7 III & $12950^{+550}_{-650}$   &  $3.39^{+0.06}_{-0.11}$      & $6.7^{+0.8}_{-1.0} $      &$3.37^{+0.02}_{-0.03} $              &$11.7^{+0.67}_{-0.5}$             & 8  & 0.04 \\ 
B331\footnote{~B163 and B331 were too embedded to allow for full modeling. Physical parameters were estimated by spectroscopic classification and estimating $T_{\rm eff}$ and $L$ from Kurucz calibration tables.} & late-B &   13000                  &  $-$                         & $13.3^{+0.9}_{-0.9}$       &$4.10^{+0.37}_{\downarrow}$         &$21.8^{+9.6 }_{-7.2}$              & 12 & 0.02  \\ 

\hline
\end{tabular}
\tablefoot{Data from Tables 2, 3 and 5 in \cite{ramirez-tannus2017}. The adopted distance for M17 is 1.98 kpc.}
\end{minipage}
\label{tab:stellar_properties}
\normalsize

\end{table*}

The data used in this work consist of near- to mid-infrared photometry ($1-11$ \micron) and the NIR part ($1-2.4$ \micron) of spectra obtained with VLT/X-shooter.  

\subsection{Spectra} \label{sec:data_spectra}
The X-shooter spectra \citep{vernet2011} of the 5 young stars studied in this paper were first analyzed by RT17, who derived stellar parameters from quantitative spectroscopy and optical ($\lambda \lesssim 1$ \micron) spectral energy distribution (SED) fitting. For a detailed listing of the spectroscopic observations used in this study we refer to that paper. Only for B275 we used a different spectrum, observed on 2019-06-06 for Program 0103.D-0099 (P.I. Ram\'irez-Tannus). In \Cref{tab:stellar_properties} we summarize the results from RT17 that we use directly in our analysis. 

For the fits of the CO bandheads we only used the NIR part (1-2.4 \micron, with spectral resolution R = 13000) of the spectra, which contain both the $1^{\rm st}$ and the $2^{\rm nd}$ overtone emission. We (re-)reduced these spectra using version 3.3.5 of the X-shooter pipeline \citep{modigliani2010}, running under the ESO Reflex environment \citep{freudling2013} version 2.11.0. The standard settings of the pipeline resulted in flux- and wavelength-calibration issues at the location of the $1^{\rm st}$ overtone bandheads. At wavelengths beyond $\sim 2.29$ \micron, the flux showed wave-like fluctuations for some objects, hindering normalization. Additionally, in this range, wavelengths were shifted by $\sim 80 \rm ~km~s^{-1}$. In our attempt to solve these problems, while at the same time obtaining an optimal telluric correction, we first reduced the objects without flux calibration and then used the {\sc  molecfit} tool \citep{smette2015a, kausch2015}, version 1.5.9, for a first telluric correction. Importantly, {\sc  molecfit} also provided a correction of the faulty wavelength calibration. We then divided the object spectra by their respective telluric standard stars that were reduced \emph{and} {\sc  molecfit} corrected in the same way. This way both the instrument response function and the bulk of the residuals of the {\sc molecfit} telluric correction are divided out. The continuum of the resulting spectra is the ratio of the continua of the science object and the telluric standard star. From this, it is in principle possible to obtain a flux calibrated spectrum by multiplying by the SED of the telluric standard star and applying some correction for slit losses. However, due to uncertainties involved in that procedure, we deemed a more reliable and consistent estimate of the continuum could be obtained from fitting SEDs to the photometric data points (\Cref{sec:phot}). Therefore, we chose to normalize the spectra and compare the observations to models that are normalized to continuum estimates obtained from fitting SEDs, as described in \Cref{sec:dust_treatment}. 

Finally, we correct for the hydrogen absorption features of the telluric standard stars in the wavelength range 1.55 - 1.8 \micron\, where they interfere with the CO $2^{\rm nd}$ overtone bandhead line fluxes. We do not correct for the stellar and disk (emission) features of the science spectra in the same wavelength range, but exclude the most contaminated regions from our fits. \footnote{A reproduction package containing the final data products can be found at \url{https://doi.org/10.5281/zenodo.7774529}}

\subsection{Photometry} \label{sec:phot}
The photometric points for our objects were taken from online catalogs and literature. We used photometry from the DENIS \citep{fouque2000}, 2MASS \citep{skrutskie2006}, Spitzer GLIMPSE \citep{reach2005} and WISE \citep{wright2010,jarrett2011} point source catalogs. For B275 and B331 we could add the extra points around 10\,\micron~ available from \citet{nielbock2001} and \citet{kassis2002}. Though a few points beyond these wavelengths were available from the same papers, we did not include them in the final fits because they likely contain envelope emission, which was not considered in our models. All adopted photometric data per object are listed in \Cref{app:phot_tables}.

\section{Methods: analytical disk model with dust and gas} \label{sec:methods}

\subsection{Keplerian disk model}\label{sec:disk_model}
The disk model that was used to fit the CO bandheads and continuum constitutes a flat disk with a Keplerian rotation profile, containing dust and/or CO gas. \footnote{A reproduction package can be found at \url{https://doi.org/10.5281/zenodo.7774529}} The disk has a radial temperature and surface density profile for the dust and the gas, described by analytical power laws as follows:
\begin{align} 
   T (r) ~ &= ~ T_i   \cdot  \left ( \frac{r}{R_i} \right ) ^p \label{eq:temp_prof} \\ 
   N_\text{H} (r) ~ &= ~ (N_\text{H})_i   \cdot  \left ( \frac{r}{R_i} \right ) ^q \hspace{3mm}  \label{eq:dens_prof}
\end{align}
where $T$ is the temperature in K, $N_\text{H}$ the hydrogen (H$_2$) surface density in cm$^{-2}$, and $T_i$, $(N_\text{H})_i$ their initial values at the inner radius $R_i$. The hydrogen surface density is converted to a $^{12}\rm CO$ gas density using the canonical CO/H$_2$ abundance of $10^{-4}$ \citep{lacy1994} and to a dust particle density using a gas to dust mass ratio of 100 \citep{bohlin1978}. 
For including \thirteenco~in the models, several abundances were tested (see \cref{sec:model_exploration}).  For the final results a standard interstellar abundance ratio $N(^{12}$CO)/$N$(\thirteenco) of 89 was used, taken from the HITRAN database\footnote{\url{https://hitran.org/docs/iso-meta/}}, which reproduces the isotopologue lines well.
To avoid confusion, all densities are reported as hydrogen column density $N_\text{H}$. 

\subsubsection*{Independent treatment of gas and dust}
\begin{table}
\footnotesize
\centering
\caption{Parameter values in grid for SED fitting.}  
\begin{minipage}{\hsize}
\centering
\renewcommand{\arraystretch}{1.4}
\setlength{\tabcolsep}{3pt}
\begin{tabular}{lcccc}
\hline\hline
parameter & symbol (unit) & min & max & step size\\
\hline
Initial temperature & $T_i$ (K)  & 200 & 1500 & 100 \\
Temperature exponent & $p$    &   $-5.0$ &    $-0.1$ &0.2  \\
Initial column density & $(N_\text{H})_{i}$ ($\text{cm}^{-2}$)  & $10^{17}$ & $10^{27}$ & $10^{x+1}$\\
Column density exponent & $q$ & $-5.6$ & $-0.1$ &  0.4  \\
Inclination & $i$ ($^{\circ}$)   & 10 & 80 & 10   \\
\hline
\end{tabular}
\end{minipage}
\label{tab:sed_grid}
\normalsize
\end{table}

\begin{table}
\footnotesize
\centering
\caption{Parameter values in grid for SED fit of B331.}  
\begin{minipage}{\hsize}
\centering
\renewcommand{\arraystretch}{1.4}
\setlength{\tabcolsep}{3pt}
\begin{tabular}{lcccc}
\hline\hline
parameter & symbol (unit) & min & max & step size \\
\hline
Initial temperature & $T_i$ (K)  & 250 & 950 & 78 \\
Temperature exponent & $p$    &  $-5.0$   & $-0.1$ & 0.2  \\
Initial column density & $(N_\text{H})_{i}$ ($\text{cm}^{-2}$)  & $10^{17}$ & $10^{27}$ & $10^{x+1}$\\
Column density exponent & $q$ & $-5.0$ & $-0.1$ &  0.4  \\
Inclination & $i$ ($^{\circ}$)   & 10 & 80 & 10   \\
Inner radius & $R_i$ (AU) & 1 & 160 & 10 \\
\hline
\end{tabular}
\end{minipage}
\label{tab:sed_grid_B331}
\normalsize
\end{table}

In all models that were used for fitting, the gas and dust properties are treated independently of each other, i.e. each have their own values. The reason for this approach stems mainly from the fact that the CO emission originates from hot gas well within the dust sublimation radius, with temperatures well above the assumed dust sublimation temperature of 1500 K. We did test a fitting approach in which dust and gas were combined and coupled with a source function $\text{S}^{\rm tot}_{\lambda}$ as follows:
\begin{align} \label{eq:S_tot}
    \text{S}^{\rm tot}_{\lambda} &=\frac{\tau^{\text{CO}}_{\lambda} * \text{S}^{\text{CO}}_{\lambda} + \tau^{\rm dust}_{\lambda} * \text{S}^{\rm dust}_{\lambda}}{\tau^{\rm tot}_{\lambda}}
\end{align}
with the total optical depth $\tau^{\rm tot}_{\lambda} = \tau^{\rm CO}_{\lambda} + \tau^{\rm dust}_{\lambda}$ the sum of the CO line and dust optical depths, the CO and dust source functions given by Planck curves according to the same temperature (so in this case $\text{S}^{\text{CO}}_{\lambda} = \text{S}^{\rm dust}_{\lambda}$) and where all quantities are a function of wavelength. This revealed that the CO emission and the dust emission responsible for the NIR-MIR excess are representative of different regions in the disk, with little or negligible overlap. Where the two do overlap the main effect of importance is the optical thickness of the dust `damping' the CO emission. However, since dust only exists at lower temperatures, where the CO emission is already significantly reduced, this effect is minimal. 

Of course, even though the CO line and dust continuum emission likely do not originate from the same region, the dust continuum still has a significant influence on the strength of the \emph{observed bandheads}: for the same absolute CO emission, a stronger continuum leads to a weaker bandhead signal relative to the continuum. Therefore, it remains important to determine and model the continuum. We opted to fit the SEDs and the CO bandheads separately with a dust-only and gas-dust disk respectively; where for the latter the dust parameters were set by the best fits obtained from the SEDs. For the fitting of the SEDs not only dust continuum, but also stellar continuum was taken into account. 

We now elaborate on the specifics of the respective modeling and fitting approaches for the continuum and CO bandhead emission.

\begin{table*}[t!]
\footnotesize
\centering
\caption{Parameter values in the grid for CO bandhead fitting.}  
\begin{minipage}{\hsize}
\centering
\renewcommand{\arraystretch}{1.4}
\setlength{\tabcolsep}{3pt}
\begin{tabular}{lcc}
\hline\hline
Parameter & Symbol (unit) & Grid values \\
\hline
Initial temperature & $T_i$ (K)  & 2000~~ 3000~~ 4000~~ 4500~~ 5000~~ 5500~~ 6000~~ 7000~~ 8000 \\
Temperature exponent &  $p$    &    $ -0.5 ~~ -0.75 ~~ -1 ~~ -2~~-3 $      \\
Initial column density & $(N_\text{H})_{i}$ ($\text{cm}^{-2}$)  & $5\times10^{23}~~~~~ (1.4~~~ 3.9)\times10^{24}~~~~~(1.1~~~3~~~8.3)\times10^{25}~~~~~(2.3~~~6.5)\times10^{26}~~~~~ (1.8~~~5) \times 10^{27}$ \\
Column density exponent & $q$    & $1~~ -0.5~~ -1~~ -1.5    $\\
Inner radius &  $R_i$ ($R_{\star}$) &  $1.1~~~ 1.6 ~~~ 2.3 ~~~ 3.3 ~~~ 4.8 ~~~ 6.9 ~~~ 10 ~~~ 14 ~~~ 21 ~~~ 30 $\\
Inclination & $i$ ($^{\circ}$)   & $ 10 ~~~ 20 ~~~ 30 ~~~ 40 ~~~ 50 ~~~ 60 ~~~ 70 ~~~ 80 $    \\
Gaussian width & $v_G$ &  1~~ 2~~ 3~~ 4~~ 5 \\
\hline
\end{tabular}
\end{minipage}
\label{tab:co_grid}
\normalsize
\end{table*}

\subsection{Continuum treatment}
\label{sec:dust_treatment}
The continuum flux has a stellar component and one that arises from thermal emission of dust near the star. The dust contribution is computed using the described analytic disk model including only dust. For the dust 0.1 $\mu$m astronomical silicate was used with opacities from \cite{laor1993}. The stellar and extinction parameters as determined previously by RT17 ($T_{\text{eff}}$, $\log g$ and $A_V$; \Cref{tab:stellar_properties}) are input in our calculations using the corresponding Kurucz models \citep{kurucz1993,castelli2004} and reddening \citep{cardelli1989} the resulting SED after adding the disk. The near- and mid-infrared photometry points are then fit by means of a grid of disk models - stellar parameters remain fixed. 

The outer radius of all the dust disk models was fixed to 500 AU, regardless of the temperature at this point. We do not expect to constrain an outer radius of the entire disk, because the photometric points available for our objects are only representative of the warm to hot dust ($T \gtrsim 150$ K) and emission from cooler dust does not change the models in the fitted wavelength ranges (1-12 \micron). 

For the inner radius $R_i$ we used a radiative equilibrium approach \citep[following][]{lamers1999} to calculate the radius at which a certain dust temperature is reached as a function of the previously determined $T_{\text{eff}}$ and $R_{\star}$ (\Cref{tab:stellar_properties}). This $R_i$ is therefore dependent on $T_i$ for each object. 
This approach results in five free parameters: $T_i$, $p$, $(N_\text{H})_i$, $q$ from \Cref{eq:temp_prof,eq:dens_prof} and the inclination $i$. The grid of parameter values used for fitting is given in \Cref{tab:sed_grid}. Only for B331 a good fit could not be obtained with $R_i$ as function of $T_i$ and stellar parameters. We therefore fitted the SED of this object using a separate grid in which also $R_i$ is a free parameter and whose values are given in \Cref{tab:sed_grid_B331}. 

\subsection{CO emission treatment}
\label{sec:CO_treatment}

\subsubsection{Combining dust and gas}
In the models used to compute the normalized CO bandheads, the dust and CO gas are both included, but do not share the same temperatures and densities, as motivated above. Instead, the dust parameters were fixed by the parameters obtained from the SED fits, and the CO gas parameters were varied over in a grid of models. Only the inclination was shared between the dust and gas disk.  

The outer radius of the gas disk model is taken to be the point where the temperature drops to 600 K. Below this temperature the gas does not substantially contribute to the bandhead emission anymore, because the excited vibrational states are not sufficiently populated. The outer radius for the dust disk is again 500 AU, as before (see \Cref{sec:dust_treatment}).

Because the CO gas excitation temperatures are typically much higher than the dust temperatures and because the inner radii for the gas-disk are so close to the star (see \Cref{tab:co_grid}) the gas and the dust do not necessarily overlap. In fact, they overlap only in the models where the CO gas temperature drops to $600~\rm K$ beyond the dust sublimation radius (calculated as described in \Cref{sec:dust_treatment}). In these cases the source function in the overlap region was calculated according to \Cref{eq:S_tot}, where the CO and dust source functions are given by Planck curves according to their respective temperatures. In all regions and cases where there is no overlap, i.e. only gas and no dust or only dust and no gas, the gas and dust emission are calculated separately and added. For the CO modeling the main significance of the dust disk is its contribution to the continuum, to which the bandheads are normalized.

\subsubsection{The gas disk and parameter grid}
The vibrational-rotational level populations of CO are assumed to be in LTE, with Einstein A coefficients and level transitions taken from \citet{li2015a}. 
The line profiles of the rotational transitions are assumed to be Gaussian with a width $v_G$. This parameter accounts for intrinsic width of the lines as well as thermal and micro-turbulent velocities. 
The intensities are calculated per ring at radius $r$, with a dependence on the ring temperature and density. They are then convolved with a Keplerian velocity profile per ring and corrected for inclination, before integrating over the radial extent of the disk. Finally, the model spectra are convolved with the instrument resolution of R = 11300 and normalized to the continuum fits presented in \Cref{sec:cont_results}, adjusted only to match the inclination of the CO disk.\footnote{Because the fitted continuum models were either optically thin or star-dominated (in the case of B331) changing the inclination did not significantly change the continuum values for the \first\, and \second\,overtone CO bandhead emission.}

As the masses of the stars are previously determined (\Cref{tab:stellar_properties}; needed for Keplerian velocities), we have 7 free parameters for our model: $T_i$,  $p$, $(N_\text{H})_i$, $q$, $R_i$ from \Cref{eq:temp_prof,eq:dens_prof}, the inclination $i$ and the Gaussian width $v_G$. To fit the data we calculated a grid of models with parameter ranges given in \Cref{tab:co_grid}. The parameter values for this final grid are not all regularly spaced and are based both on literature and on several other exploratory grids that were calculated to investigate the relevant parameter space for our objects. The inner gaseous disk radius $R_i$ is given in units of stellar radii (\Cref{tab:stellar_properties}, RT17). Both the initial hydrogen column density $(N_\text{H})_i$ and the $R_i$ values are evenly spaced on a log scale. The positive value ($+1$) for the density exponent $q$ was included, because of evidence for an inverse density profile of the innermost regions of the disk \citep{antonellini2020}.

\subsubsection{Correcting for pseudo-continuum}

When calculating the final CO bandhead spectrum, each subsequent bandhead flux is added to the P-branch ($\Delta J = -1$) and higher R-branch ($\Delta J = + 1$) rotational  transitions of the previous one. This, when convolved with the instrument resolution, creates a kind of secondary continuum and influences the overall appearance of the bandhead spectrum. This is true for both overtones (see also \Cref{fig:comparison}). However, when normalizing the observed spectra in the \second\, overtone wavelength regions it is impossible to recognize such effects, especially because the overlapping line wings of the hydrogen Brackett series also influences the continuum there. To account for this pseudo-continuum effect, we introduced a correction of the bandhead continuum of the modeled $2^{\rm nd}$ overtone bandheads before fitting. We did this by fitting a $3^{\rm rd}$ degree spline through the minima just before the bandhead-onset wavelengths and dividing the modeled bandheads (already normalized to star and dust continuum) by this `bandhead continuum'. The normalization of the data in the $1^{\rm st}$ overtone region is much less problematic and the models there did not require similar corrections.

\subsection{Model exploration}
\label{sec:model_exploration}
\begin{figure*}[p!]
\vspace{-0.5cm}
\hspace{-1cm}
   \includegraphics[width=1.08\hsize]{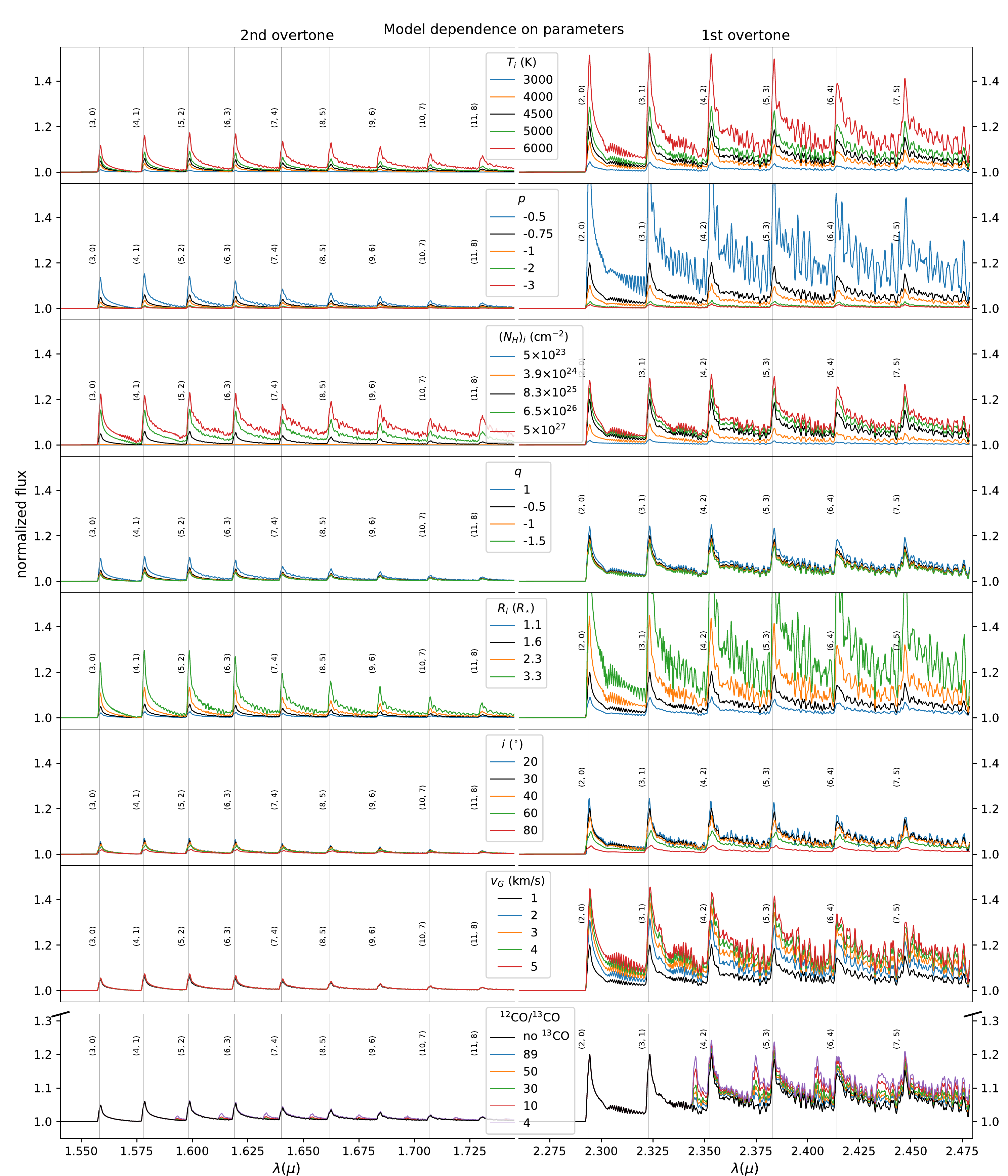}
      \caption{Overplotted models, showing for each parameter the effect of varying its value, while keeping all other parameters fixed at: $T_i$ = 4500 K; $p$ = -0.75; $(N_{\rm H})_i = 8.3 \times 10^{25}~\text{cm}^{-2}$; $q$ = -0.5; $R_i$ = 1.6 $R_\star$; $i$ = 30$^{\circ}$; $v_G = 1~\rm km~s^{-1}$. The models are all without inclusion of \thirteenco, except in the bottom panel, where the effect of adding \thirteenco~in varying abundances is shown. The continuum and stellar parameters used for these models are those of B275. The `fixed' model is colored black on each plot and is a fit to the B275 spectrum.}
         \label{fig:comparison}
\end{figure*}

\begin{figure*}[h!]
   \includegraphics[width=1.\hsize]{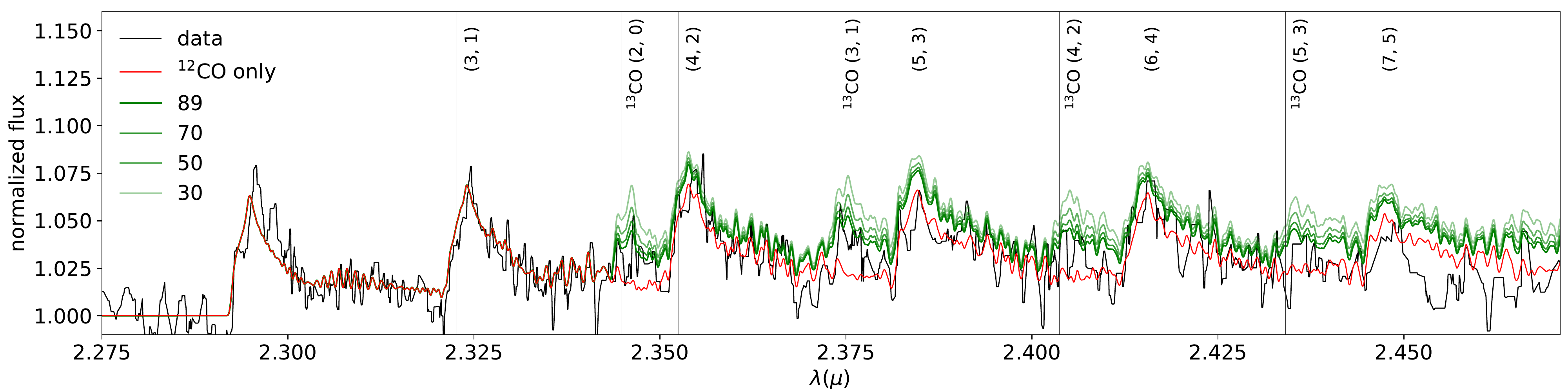}
      \caption{The observed \first~overtone CO bandheads of B163 (black line) shown together with models varying in $^{12}$CO/\thirteenco~ratio - a decreasing ratio signifying \emph{more} \thirteenco. The red line indicates the best fit model for this target, with no \thirteenco~included, the green lines indicate models differing only in the amount of \thirteenco. Including \thirteenco~with the ISM value (89) clearly improves the fit with respect to fitting $^{12}$CO only. Decreasing the ratio leads to progressively worse fits, indicating that \thirteenco~ is likely not enhanced in the circumstellar environment of this object.}
         \label{fig:B163_with_13CO}
\end{figure*}

Several examples of CO bandhead fitting using analytical disk models are available in literature \citep[e.g.][]{kraus2000,bik2004,ilee2013}. Similar to our approach, these works adopt an LTE treatment of the gas, Keplerian rotation and a 1D thin disk approximation. Here we list changes and improvements in our approach relative to these earlier efforts.

First, we introduce an internally consistent treatment of the continuum and normalization. In previous work, the bandheads are usually continuum subtracted and normalized to the first bandhead. Instead, we estimate the continuum from SED fitting and use this to normalize our models. With this approach the information contained in the strength of the bandheads is retained.

Second, we fit the first 5 bandheads of the first vibrational overtone ($v$ = 2-0, 3-1, 4-2, 5-3, 6-4) at the same time. We also predict the $6^{\rm th}$ bandhead ($v =$ 7-5), but because it is at the very edge of the observed spectral range, the signal to noise ratio is insufficient to include it in the fits.

Third, for the first time, the $2^{\rm nd}$ vibrational overtone bandheads ($\Delta v = 3$, from 1.5 \micron~ onward) are included.

Fourth, we include $^{13}$CO. These lines are weak and their diagnostic value in constraining disk properties is limited. However, including these lines allows to constrain the isotopologue ratio $^{13}$CO/$^{12}$CO. 

Finally, we probe higher temperatures than the commonly assumed dissociation temperature of 5000 K \citep{bosman2019}; grid values run up to 8000 K. Such high temperatures should be taken with care as they may not be physical, since CO dissociation is not included in the models. Allowing for higher temperatures may mimic limitations in the LTE assumption, which can overestimate the temperature if  
non-collisional excitation mechanisms, such as UV pumping, contribute to setting the state of the gas. The maximum temperatures retrieved from our fits can then serve as an additional probe of the validity of the LTE assumption.
\newline

To facilitate a comparison with previous results of similar fitting efforts in literature we identify where and how these novel aspects in our approach might impact the parameter estimation. To do this we explore the effects of the free parameters on the appearance of the 1$^{\rm st}$ and 2$^{\rm nd}$ overtone bandhead emission using \Cref{fig:comparison}. This figure was made by over-plotting several models, sequentially varying only one parameter and keeping all others fixed at a value fitting to the B275 bandheads. The fitted model is plotted in black in each panel to give an impression of where the data for this object are. In the very last panel we show the effects of including \thirteenco~in varying abundances; we discuss the inclusion of \thirteenco~at the very end of this section.

 The amount of emission in both the $1^{\rm st}$ and $2^{\rm nd}$ overtone bandheads is most sensitive to the temperature and the extent of the (hot) gaseous disk, as can been seen on the $T_i$, $p$ and $R_i$ plots in \Cref{fig:comparison}. These parameters have very similar effects on the amount of flux of the bandheads.   Since the temperature always drops going outward, a small initial radius will lead to a small area of the hottest gas, i.e. less flux. This can be compensated for by a shallower temperature drop (smaller $|p|$) resulting in a larger emitting surface. Note the strong emission in the $T = 6000~\rm K$ model, which is stronger than the commonly observed strength of these features. Again, CO dissociation is not included in our models.
 
Since the disk is assumed to be Keplerian, the velocities are set by the mass of the central star (kept constant for each object), the inner radius $R_i$  and the inclination $i$ \footnote{The line broadening velocities $v_G$ are too small to contribute observably.}. These parameters, therefore, determine the onset wavelengths of the bandheads and influence their shape. High (more edge on) inclinations and smaller inner radii broaden the bandhead profile and shift the onset wavelength blue-ward. Both these parameters, however, also influence the total flux.

For the \emph{relative} strength of the $1^{\rm st}$ and $2^{\rm nd}$ overtone bandheads optical depth effects play an important role. The lines of the 2$^{\rm nd}$ overtone are intrinsically weaker and therefore remain optically thin for higher column densities. For this reason line broadening, i.e. higher $v_G$, leads to higher flux only for the 1$^{\rm st}$ overtone bandheads and not for those of the 2$^{\rm nd}$ overtone. The inclination has slightly less effect on the 2$^{\rm nd}$ overtone flux than on the $1^{\rm st}$ overtone flux: the radiating surface of an optically thick inclined disk is reduced, but the $2^{\rm nd}$ overtone remains optically thin for longer sight-lines through the disk. 

Finally, increasing the initial column density $(N_\text{H})_i$, or decreasing $|q|$, beyond certain grid values has a larger impact on the 2$^{\rm nd}$ overtone flux, because these lines remain optically thin within the grid values. These changes to the \second\, overtone flux are more detectable than the changes to the \first\, overtone within crucial ranges of $(N_\text{H})_i$. Therefore, including the 2$^{\rm nd}$ overtones in the fits especially helps towards constraining the column density. 

We now discuss the effects of each of the four modeling improvements.

\subsubsection{Continuum normalization}

It is important to keep in mind that the stellar parameters and thermal dust continuum that are kept fixed for each star in the fitting process, do strongly influence the appearance of the bandheads - and therefore the fit results. A thorough discussion of these effects is given in \Cref{app:impact_stellar_params}. Here we mention the most important effect: a higher continuum originating from the star and dust disk will lead to weaker bandheads for the same gas disk model.  \\
The most important constraint from including the bandhead strength relative to the continuum is on the ranges of values for $p$ and $v_G$. Shallow temperature gradients and high intrinsic line velocities lead to very high bandhead fluxes beyond observed ranges, that cannot be compensated for by other parameters. Thus, we limited the probed ranges in our grid accordingly (\Cref{tab:co_grid}).   

\subsubsection{Fitting multiple $1^{\rm st}$ overtone bandheads}
The effect described earlier (in \Cref{sec:CO_treatment}) that the bandhead series, if strong enough, creates a secondary continuum, can be clearly seen on \Cref{fig:comparison} for any parameter that influences the strength of the bandheads. The later bandheads (i.e. those with higher vibrational quantum numbers) are also more sensitive to temperature changes, likely due to excitation effects. 
These effects generally lead to better temperature constraints when fitting more bandheads together.

\subsubsection{Including the $2^{\rm nd}$ overtone bandheads}

Following from our previous observations, the most important determinant of the detectability and (relative) strength of $2^{\rm nd}$ overtone bandheads is the column density (profile) of the disk. We tested this by comparing fit results with or without including $2^{\rm nd}$ overtones. The exponent $q$ is poorly constrained in general and even its selective effect on the \second\,overtone did not change fit results significantly. However, including the \second\, overtone did yield significant changes for the initial column density $(N_\text{H})_i$.

\subsubsection{Including $^{13}$CO}

CO bandhead emission is a signature not unique to YSOs - it is evidence for the presence of a (hot and dense gaseous) disk, but as such it is also observed to originate in the (decretion) disks of evolved B[e] stars \citep{kraus2000}. In these stars the stellar surface layers can be enriched in $^{13}$C, due to the evolution of the (massive) star and chemical mixing processes. Stellar winds then cause this anomalous isotopologue ratio to be carried into the circumstellar environment, where consequently the N($^{12}$CO)/N(\thirteenco) abundance ratio drops and \thirteenco\,bandhead signals are enhanced. \cite{kraus2020a} show that therefore, including \thirteenco~emission in the models, can provide a test to distinguish the evolved objects from the younger ones.

\begin{table*}[htb!]
\footnotesize
\caption{Parameter ranges for which reasonable SED fits could be obtained to the available photometry.}        
\centering
\renewcommand{\arraystretch}{1.4}
\begin{tabular}{c|cc|cc|cc|cc|cc}
\hline
\multicolumn{1}{c|}{Object  } &\multicolumn{2}{c|}{$T_i$ (K) \hspace{2pt} [$R_i$ (AU)]} & \multicolumn{2}{c|}{$-p$} & \multicolumn{2}{c|}{$(N_\text{H})_{i}$ ($\text{cm}^{-2}$)}   & \multicolumn{2}{c|}{$-q$} & \multicolumn{2}{c}{$i$ ($^{\circ}$)} \\
~ & best fit & range   & best fit & range & best fit & range & best fit & range & best fit & range  \\
\hline                                                                    
B163 & 1500 [1.4]\tablefootmark{a} & 1300 - 1500 [1.8 - 1.4]\tablefootmark{a} & $0.1$ & 0.1 - 0.5 & $10^{22}$ & $10^{21} - 10^{22}$ & $4.0$ & $3.0$ - $4.8$ & 80 & 10 - 80 \\
B243 & 1400 [5.8]\tablefootmark{a} & 1200 - 1500 [7.7 - 5.0]\tablefootmark{a} & $0.7$ & 0.1 - 2.1 & $10^{21}$ & $10^{20} - 10^{21}$ & $5.6$ & 0.1 - 5.6 & 80 & 10 - 80 \\
B268 & 1500 [4.7]\tablefootmark{a} & 1300 - 1500 [6.1 - 4.7]\tablefootmark{a} & $0.7$ & 0.1 - 2.5 & $10^{21}$ & $10^{20} - 10^{21}$ & $5.4$ & 2.3 - 5.6 & 80 & 10 - 80 \\
B275 & 1500 [7.8]\tablefootmark{a} & 1200 - 1500 [12.0 - 7.8]\tablefootmark{a} & $0.9$ & 0.1 -1.5 & $10^{21}$ & $10^{19} - 10^{22}$ & $2.3$ & 0.1 - 5.6 & 60 & 10 - 80 \\
B331 & 412 ($R_i = 31$)\tablefootmark{b} & 400 - 700 ($R_i = 11 \text{-} 71 $)\tablefootmark{b} & $5.0$ & 0.7 - 5 & $10^{23}$ & $10^{23} - 10^{27}$ & $0.1$ & 0.1 - 5.0 & 40 & 10 - 80 \\

\hline
\end{tabular}
\label{tab:sed fit ranges}
\begin{itemize}[label={}]
\item $^{(a)}$ The inner radius $R_i$ is reported in square brackets next to the initial temperature $T_i$ as the distance from the star at which this temperature is reached.
\item $^{(b)}$ Only for B331 $R_i$ is a free parameter varied over in a grid.
\end{itemize}
\end{table*}

\begin{table}[htb!]
\footnotesize
\centering
\caption{Properties of the inner dusty disk.}        
\begin{minipage}{0.9\hsize}
\centering
\renewcommand{\arraystretch}{1.4}
\setlength{\tabcolsep}{3pt}
\begin{tabular}{lcccccc}
\hline
Object & $R_i$ & $R_{\text{out}}$ & $T_{\text{out}}$ & $(N_\text{H})_{\text{out}}$ & area (AU$^2$) & $M_{\text{tot}}$ (\Msun)\footnote{The mass is the total gas and dust mass of the ring, under the assumption of a gas to dust mass ratio of 100.} \\
\hline
B163 & 1.4 & 11.9 & 1208 & $1.6\times 10^{18}$ & 881 & $2.2\times 10^{-8}$ \\
B243 & 5.8 & 13.5 & 771 &$ 8.4\times 10^{18}$ & 936 & $2.1\times 10^{-8}$ \\
B268 & 4.7 & 11.1 & 816 & $9.1\times 10^{18}$ & 638 & $1.4\times 10^{-8}$ \\
B275 & 7.8 & 30.4 & 442 & $4.6\times 10^{19}$ & 5437 & $1.7\times 10^{-7} $\\
B331 & 31 & 37.2 & 166 & $9.8\times 10^{22}$ & 1847 & $\geq 5\times 10^{-5}$ \\

\hline
\end{tabular}
\tablefoot{\textit{The inner dusty disk} is defined as the ring of dust where $99\%$ of the modelled 1-12 \micron~dust emission comes from. Where not provided, units are the same as in \cref{tab:sed fit ranges}. }
\end{minipage}
\label{tab:dust_disk_props}
\normalsize
\end{table}

The objects in our sample are expected to be YSOs due to their young cluster environment. To assess the sensitivity of our results to the \thirteenco~feature, we computed models for gradually increasing \thirteenco~abundances. In the last panel of \Cref{fig:comparison} we show the model with only $^{12}$CO and with ratios of $^{12}$CO/\thirteenco~down to 4 (the $\sim 22$-fold enhancement of the abundance found by \cite{kraus2020a} in a B[e] supergiant). Even for the standard (interstellar) ratio 89, \thirteenco~ leaves a detectable signal in the \first~overtone, which becomes very prominent for the highest abundances. In \Cref{fig:B163_with_13CO} we show the observed \first~overtone bandheads of B163, the object for which the \thirteenco~feature is most pronounced, together with its best fit model and models with varying \thirteenco~abundance. When included with the standard abundance, the \thirteenco~feature clearly improves the fit; but increasing the abundance leads to progressively poorer results, limiting a possible enhancement of the abundance to $\sim 30\%$ in this case. We also find that including \thirteenco~does not change the other disk parameters that are fitted. These findings are similar in the other objects; save for B331, where the hydrogen Pfund line series interferes with the \thirteenco\,bandhead signal and we cannot assess the impact of the isotopologue. We conclude that, although the inclusion of \thirteenco~is of limited diagnostic value to constrain disk parameters, its detected features, consistent with an interstellar abundance, do provide a confirmation of the young nature of our objects. Therefore, when presenting and plotting our results we use the models including \thirteenco, except for B331. 

\begin{figure*}[htb!]
            \includegraphics[width=\hsize]{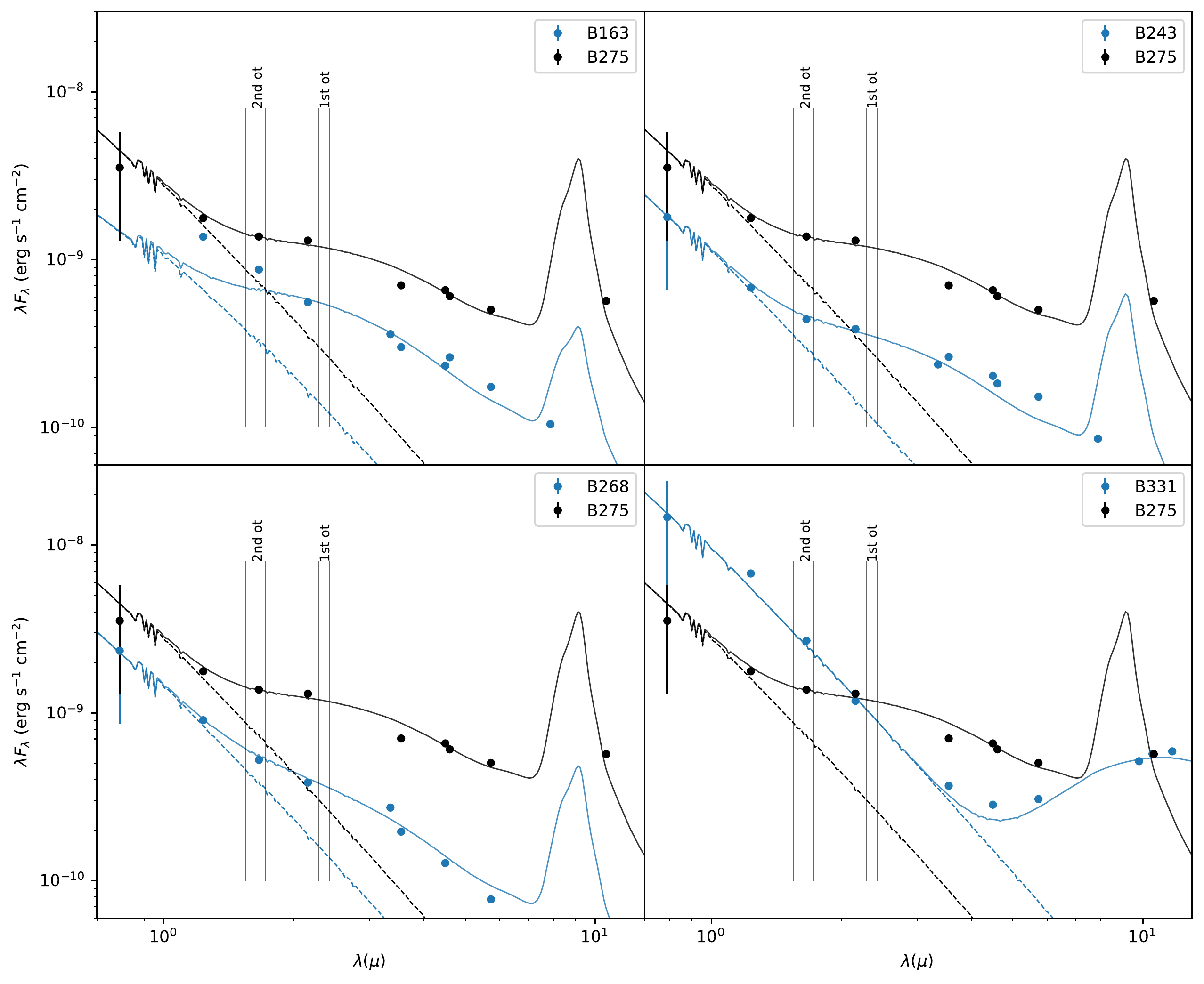}
      \caption{Best fit model SEDs and photometric data points. On each figure the data and results for B275 (in black) are plotted along with one other object (in blue) for comparison. Kurucz models representing the stellar spectrum are given in dashed lines; the combined stellar and disk model in solid lines. The scattered points mark the photometry. Also indicated are the wavelength ranges of the 2$^{\rm nd}$ and 1$^{\rm st}$ overtone bandheads. Note how B331 is both more luminous and also lacks NIR excess, with the dust emission starting around 4 to 5 \micron.}
         \label{fig:sedfits}
\end{figure*}
\section{Results} \label{sec:results}

\subsection{Continuum fits} \label{sec:cont_results}
 
The aim of the SED fits is, first, to determine a reliable continuum for normalizing the modeled CO bandheads, and second, to obtain constraints on the inner parts of the dust disk.

\Cref{fig:sedfits} visualizes the best fitting model SEDs for the five objects in our study, overplotted with the available photometric points. The case of B331 is different and we discuss it separately. 

For the objects B163, B243, B268 and B275 the available photometry could be fitted well with optically thin dust emission. One of the consequences of this is that the inclinations remain unconstrained. We used a reduced $\chi^2$ statistic, in addition to an inspection by eye, to determine the range over which a reasonable fit can be obtained for each parameter. These ranges, along with the best fits are reported in \Cref{tab:sed fit ranges}. The temperature $T_i$ and column density $(N_\text{H})_i$ at the inner rim are reasonably well constrained, with $T_i \approx 1500$ K and $(N_\text{H})_i \approx 10^{21}$  $\rm cm^{-2}$ for all four objects. The temperature and density exponents $p$ and $q$ are degenerate: a less extended emission region (higher $|q|$) can be compensated for with a slower temperature decline (lower $|p|$). The derived values of the exponents are different than for the gaseous disk (\Cref{sec:CO_results}), with the values for $p$ generally shallower and the values for $q$ steeper than those obtained by fitting the CO bandheads. Apart from degeneracies, this may also reflect the difference between the disk regions probed, i.e. the gaseous part close to the star (see \cref{sec:CO_results}) versus the inner parts of the dust disk. 

With respect to the best fit model, B163 has an excess in the J ($\sim 1.2~\mu \rm m$) and the H ($\sim 1.7~\mu \rm m$) bands. This could be because the disk is not actually optically thin (see also \Cref{sec:limitations}) or due to refractory dust grains with sublimation temperatures in excess of 1500 K. The normalization of the \second\,overtone bandheads is only slightly affected; maximally by a factor of 1.3. 

The SED of B331 could not be fit with the grid used for the other objects. It is the only object which does not have a NIR dust excess within the X-shooter spectral range (the excess begins at $\lambda \sim 3.6~\mu$m), signifying that the star is surrounded by cooler dust. Here too, the parameter that is best constrained is the dust temperature at the inner rim. However, when using the radiative equilibrium approach to calculate $R_i$ the cool dust disk would start as far as $\sim 140$ AU, leading to poor fits. We opt to fit B331 with $R_i$ as a free parameter and arrive at a dust disk with $R_i \approx 30~ \rm AU$. For this object the modeled emission is optically thick, hence the best fit column density should be considered a lower limit. The inclination is again poorly constrained, this time because it is degenerate with the density decline $|q|$, i.e. extent of the disk.

We use these best fits and their parameters when modeling the dust for the CO bandhead models and for normalizing those models, as described in \Cref{sec:CO_treatment}. It is more challenging to use the results to truly constrain the parameters of the inner dust disk. 
Ultimately, because the data are simply too scarce to determine all the parameters, we find it more realistic to describe the results in terms of an amount (in the optically thin case) or surface (in the optically thick case) of dust in a ring around the star at a certain temperature. Because our photometry points only reach up to 12 \micron, they are only representative of dust emission from a limited part of the dust disk, i.e. the hotter, inner parts. To understand and compare the characteristics of these parts we define the inner dust disk as the part of the disk where $99 \%$ of the dust emission between 1-12 \micron~comes from. Using the best fitting model we can then provide an inner and outer radius and a surface area for this emission region, and determine the temperatures and column densities at those radii. For B331 we calculate the inclined disk surface, because of the emission being optically thick. Integrating the column density between inner and outer radii provides a total mass for the emission region. The results are reported in \Cref{tab:dust_disk_props}. 

The inner and outer rim column densities as well as the masses of the defined emission region are very similar for the first three objects, even though their outer temperatures ($T_{\rm out}$) and their areas are rather different. In combination with identical inner rim temperatures ($T_i$), it follows that the best fitting models are not sensitive to the size of the emission region and its temperature structure, and that it is thus the hottest dust that dominates the emission. B275 clearly has more dust emission, which is reflected in a higher disk mass.   

The dust emission from B331 is dominated by longer wavelength emission from cooler dust. The NIR emission seen in the other objects is absent (\Cref{fig:sedfits}). The fact that the inner rim temperature is so low suggests a dust-free inner cavity in the disk. A higher density and a slower density decline lead to a much higher disk mass than for the other objects. The caveat here is, however, that it is possible that the three longest wavelength points are contaminated with envelope emission. The even longer $20~\mu \rm m$ and $37~\mu \rm m$ points from \cite{lim2020} could not be fit with our disk model, most likely due to the presence of envelope emission which we did not include in our models. 

\subsection{CO bandhead fits} \label{sec:CO_results}
\begin{table*}[ht!]
\footnotesize
\centering
\caption{Fit results for CO bandhead modeling (see text for details).} 
\begin{minipage}{0.83\hsize}
\centering
\renewcommand{\arraystretch}{1.4}
\setlength{\tabcolsep}{3pt}
\begin{tabular}{l|cc|c|cc|c|cc|cc|c}
\hline\hline
Object &   & $T_i$ (K)   & $p$ &  & $(N_\text{H})_i$ ($\text{cm}^{-2}$)   & $q$   &  & $R_i$ ($R_*$)  &  & $i$  ($^{\circ}$)  & $v_G$  \\
    & max  prob  & bf &   &  max prob  & bf   &  &  mean   & bf & max prob  & bf & ($\rm km~s^{-1}$ ) \\
\hline
B163 &  $5000^{+1700}_{-1700}$ &  4000 &  -3 &  ${8.3^{+96}_{-7.7}\times 10^{25}}$ &  $8.3\times 10^{25}$ &  -1.5 &    $3^{+2.7}_{-1.4}$ &  3.3 &   $48^{+20}_{-20}$ (mean) &  50 &  1 \\
B243 &  $2000^{+3000}$ &  2000 &  -3 &  ${1.1^{+15}_{-1}\times 10^{25}}$ &   $1.1\times 10^{25} $&  -1.5 &   $3.8^{+4.2}_{-2}$ &  14 &   $45^{+20}_{-20}$ (mean) &  50 &  1 \\
B268  &  $4500^{+1000}_{-1000}$  &  4500 &  -0.75 &  ${3^{+5.8}_{-2}\times 10^{25}}$  &  $3\times 10^{25}$ &  -1.5 &   $1.8^{+0.73}_{-0.52}$ &  2.3 &  $80_{-22}$  &  80 &  2 \\
B275  &  $3000^{+2500}_{-1000}$ &   4000 &  -2 &  ${3^{+22}_{-2.6}\times 10^{25}}$ &   $3\times 10^{25} $&  -1.5 &   $4.7^{+4.1}_{-2.2}$ &  4.8 &  $30^{+26}_{-20}$ &  30 &  1 \\
B331\footnote{B331 was fitted without including \thirteenco.} &  $3000^{+2500}_{-1000}$ &  3000 &  -3 &  ${1.1^{+17}_{-1}\times 10^{25}}$ & $ 1.1\times 10^{25}$ &  -1.5 &    $3.2^{+3.2}_{-1.6}$ &  4.8 &  $20^{+32}_{-10}$ &   30 &  1 \\
\hline
\end{tabular}
\tablefoot{The parameters $p$, $q$ and $i$ were fixed to their maximum probability values. For the other parameters either the mean or maximum probability (max prob) value are quoted, along with the best fitting (bf) model parameters.}
\end{minipage}
\label{tab:co_best_fits_13CO}
\normalsize
\end{table*}

\begin{table}[h]
\footnotesize
\centering
\caption{Properties of the CO emission region.}   
\begin{minipage}{\hsize}
\centering
\renewcommand{\arraystretch}{1.4}
\setlength{\tabcolsep}{2pt}
\begin{tabular}{lccccccc}
\hline
Object & $R_i$ & $R_{\text{out}}$ & $R_{\text{dust}}$\footnote{The inner radius of the dust disk.} & $(N_\text{H})_{R_i}$\footnote{Column density at $R_i$.}  & $(N_\text{H})_{R_{\text{out}}}$\footnote{Extrapolated column density at $R_{\text{out}}$.} & area (AU$^2$) & $M_{\text{tot}}$\footnote{The mass $M_{\text{tot}}$ is the total gas mass of the CO emitting region, based on the assumed CO abundance $\rm CO/H_2=10^{-4}$.} (\Msun) \\
\hline
B163  &  0.16  &  0.29  &  1.4  & 8.3 $\times 10^{25}$ &  3.2$\times 10^{25}$ & 0.39  &  3.6$\times 10^{-6}$ \\
B243  &  0.5  &  0.75  &  5.8   & 1.1 $\times 10^{25}$ &  5.9$\times 10^{24}$ & 1.9  &  2.9$\times 10^{-6}$ \\
B268  &  0.094  &  1.4  &  4.7  & 3 $\times 10^{25}$ &  5.3$\times 10^{23}$ & 12  &  3.6$\times 10^{-6}$ \\
B275  &  0.26  &  0.67  &  7.8  & 3 $\times 10^{25}$ &  7.2$\times 10^{24}$ & 2.4  &  5.9$\times 10^{-6}$ \\
B331  &  0.48  &  0.83  &  31   & 1.1 $\times 10^{25}$ &  4.8$\times 10^{24}$ & 2.8  &  3.7$\times 10^{-6}$ \\
\hline
\end{tabular}
\tablefoot{\textit{The CO emission region} is defined as the part of the disk (bounded by $R_i$ and $R_{\rm out}$) where the best fitting model CO emission originates. All radii are given in AU and column densities in cm$^{-2}$. }
\end{minipage}
\label{tab:co_disk_props}
\normalsize
\end{table}

\begin{sidewaysfigure*}[p!]

   \includegraphics[width=1.0\textwidth]{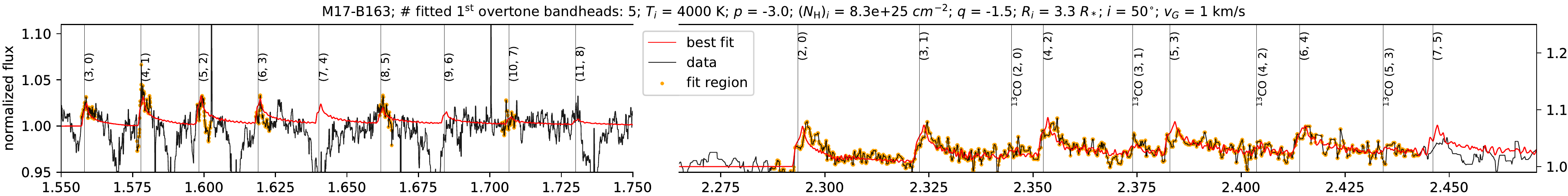}
    
   \includegraphics[width=1.0\textwidth]{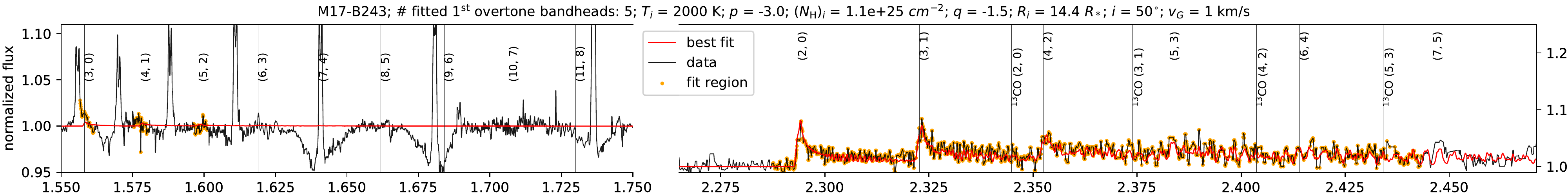}
     
   \includegraphics[width=1.0\textwidth]{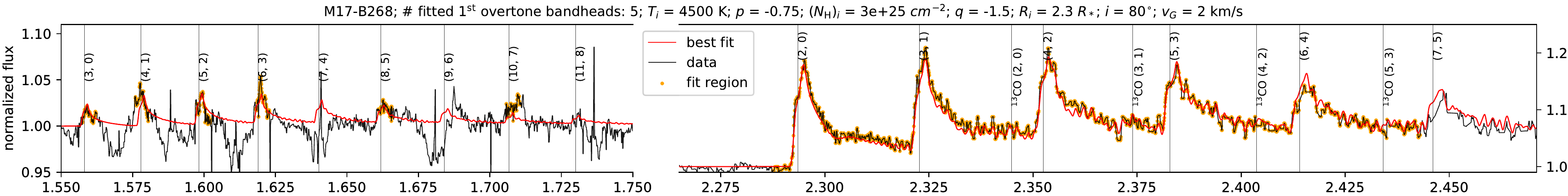}
    
   \includegraphics[width=1.0\textwidth]{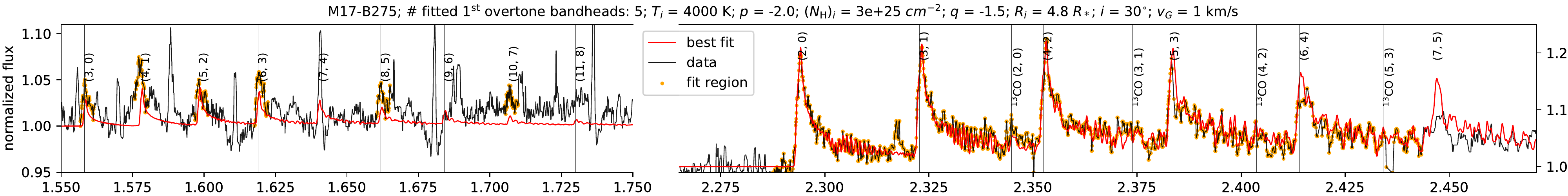}
     
   \includegraphics[width=1.0\textwidth]{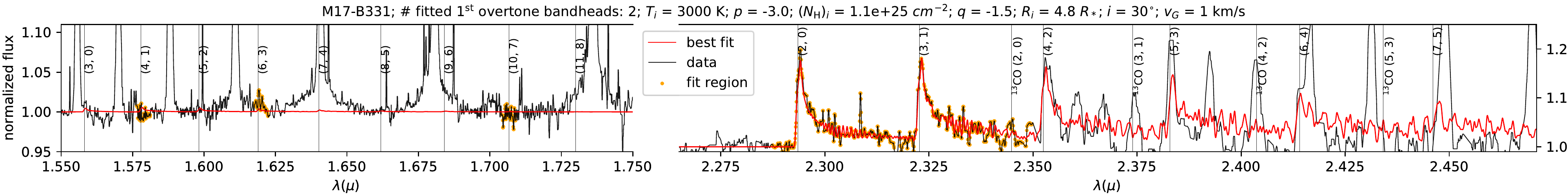}

         \caption{Best fits (in red) overplotted with data (in black) for all objects. The data points included in the fits are marked in orange. Note the different y-axis scales for the \first\,(right) and \second\, (left) overtone. All fluxes are normalized. On the \second\, overtone bandhead plots the hydrogen Brackett series (not marked) can also be seen, in absorption from the stellar photosphere and/or in emission from the disk.}
         \label{fig:co_fit_all}
\end{sidewaysfigure*}

The best fits resulting from the grid of models described in \Cref{sec:CO_treatment}, are plotted for each object in \Cref{fig:co_fit_all}, together with the data and the fit regions. For B331 we fitted only two of the $1^{\rm st}$ overtone bandheads, due to strong emission in the hydrogen Pfund line series. All other spectra are fitted including five $1^{\rm st}$ overtone bandheads. The $2^{\rm nd}$ overtone bandheads suffer from poor signal to noise due to a combination of hydrogen and other emission features and telluric contamination. We fitted only those bandheads that were more or less free from hydrogen emission. Even if they were not clearly detected, at least some $2^{\rm nd}$ overtone bandhead regions were always included in each fit, as their absence also provides constraints. 

Because of the degeneracy between different parameters and limits to the data quality and spectral resolution, it was non-trivial to understand and determine the accuracy of fits to the data. The most straightforward approach of quoting the minimal $\chi^2$ best fit values was insufficient as `next best' fits sometimes produced very different parameter values. In addition, especially the power law exponents $p$ and $q$ were poorly constrained within the probed ranges. To get better insight into the parameter space and to determine errors, we  used marginalized likelihood distributions for each parameter, as described in detail in \Cref{app:fits_errors}.  One caveat with this approach is that, in general, the errors and mean depend on the parameter space chosen. In this case, the parameter space is large and limited to physically reasonable values, so that we can assume that likelihoods for parameter values outside the grid are (close to) zero. To better understand the temperature, density and velocity information, we fixed the  $p$, $q$ and $v_G$ parameters at their maximum likelihood values after fitting with the entire grid. We determined the best fits for the remaining parameters ($T_i$, $(N_{\text{H}})_i$, $R_i$, $i$) from a `reduced' grid with the $p$, $q$ and $v_G$ parameters fixed. 

The results are summarized in \Cref{tab:co_best_fits_13CO}. For $T_i$, $(N_{\text{H}})_i$, $R_i$ and $i$ we determined both the most probable and the mean values, with their respective uncertainties, based on the marginalized probability distributions. As explained in \Cref{app:fits_errors}, the best fit, the most probable and the mean values can be rather different, but for almost all parameters the best fit values fall close to and within the error bars of either the mean or the most probable value or both - we list the one that is closest to the best fit value. In the table the most probable value is indicated as `max prob', the mean value as 'mean' and the best fit as `bf'. Only for the inner radius $R_i$ of B243 does the best fit value in the reduced grid lie outside the error bars of the quoted mean. This is the object with the weakest bandheads and the poorest signal to noise.

In order to illustrate and compare these results in \Cref{tab:co_best_fits_13CO} a bit better we made a similar emission region analysis as for the inner dust disk (\Cref{tab:dust_disk_props}). In \Cref{tab:co_disk_props} we show the inner and outer radii of the part of the disk where, according to the best fitting model, the CO emission originates. By construction the outer radius $R_{\text{out}}$ is given by the point at which the gas temperature drops to $600~\rm K$. We also show where the dust disk begins ($R_{\text{dust}}$) and provide the column densities at $R_i$ and $R_{\text{dust}}$. Finally, we provide the total gas mass of the CO emitting region (since there is no dust there), based on the assumed CO abundance. 

\subsection{Summary of the most important aspects of the results}

\begin{figure*}
\centering
   \includegraphics[width=1.0\hsize]{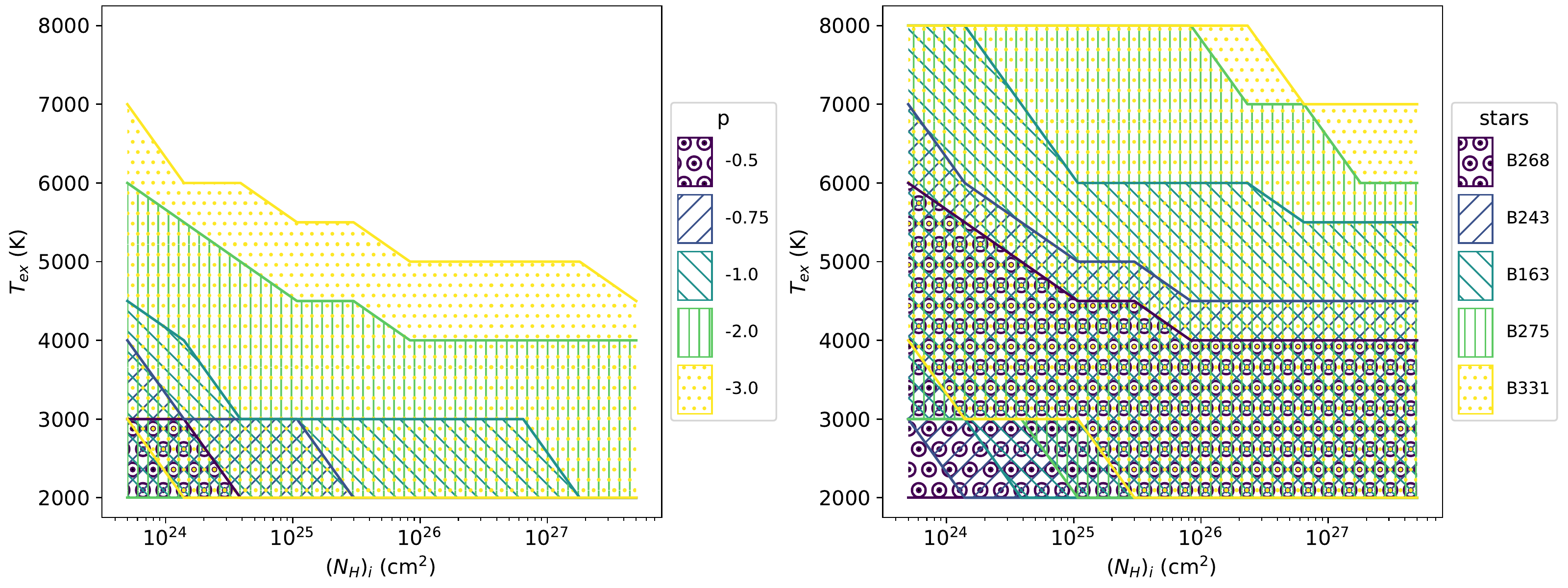}
      \caption{Illustration of the temperature and column density ranges that, for fixed parameters $R_i \approx 0.25$ AU, $q=-1.5$, $i=40^{\circ}$ and $v_G=1~ \rm km~s^{-1}$, lead to CO \first\, overtone bandhead maximum flux values between 2 and 25\% continuum, which represent the minimum detectable and the maximum observed normalized fluxes respectively. Left: the accessible parameter space for different values of the temperature exponent $p$ (models are for B268). Right: the same, for the different stars as varying in mass and continuum emission (models are for $p=-2$).} 
      
     \vspace{1pt}
         \label{fig:param_space_corner}
\centering
\end{figure*}

Before we discuss our results we briefly bring together the different aspects that can help form a picture of the studied disks. Firstly, the SEDs are fit with a thin dusty disk, but the parameters remain poorly constrained apart from the temperature at the inner rim. Except for B331 all inner rim temperatures are at or very close to the assumed dust sublimation temperature, suggesting that these inner dust disks are undisrupted. B331 is an exception with cooler dust at larger radii, pointing to a perturbed inner dust disk (see \Cref{sec:kindofdisk} for further discussion). The SED-derived densities we will leave aside for reasons explained in \Cref{sec:limitations}.

From \Cref{tab:co_best_fits_13CO,tab:co_disk_props} we see that in all cases the CO bandheads originate in a relatively narrow ring close to the star. There does not appear to be a relation between $R_i$ and $T_i$. Temperatures range between 2000 and 5000 K - the lower limit given by our grid, the upper likely being related to the temperatures at which CO abundances become negligible due to hot gas chemistry dissociating the CO molecules \citep{bosman2019}, the local radiation field temperature being too low to cause photo-dissociation. Thus, despite probing higher temperatures we retrieve the expected maxima in the best fit values, confirming that even if non-LTE mechanisms are at work, it is unlikely that they would drastically change our results. Taking into account the continuum and fitting multiple bandheads are important measures to ensure that the emission region and temperatures are reliably constrained.

The distribution of best-fit inclinations is consistent with random orientations. The Gaussian line widths are in the order of the thermal velocities at the CO temperatures, suggesting that turbulent or other stochastic motions are not significant in the emitting regions. 

Though the column densities are similar for all objects, we see that they are slightly lower (by a factor of $\sim 3-8$) where the $2^{\rm nd}$ overtone is not observed (see objects B243, B331 in \Cref{fig:co_fit_all}). This is consistent with the \second\,overtone being observed only when column densities are high enough. The fact that the emission originates from a narrow ring means that the density change over larger parts of the disk and hence $q$ will be poorly constrained (also apparent on \Cref{fig:comparison}). Nonetheless, the most likely value for $q$ was found to be identical for all objects. The last column in \Cref{tab:co_disk_props} shows that the total masses of the CO emitting region are surprisingly similar. This largely relates to, but is not fully explained by, the similarity in the derived densities. 

In general we note that, given the accessible parameter space, the overall similarity in the emitting regions is striking. To illustrate this we show in \Cref{fig:param_space_corner} the range of temperatures and column densities that are in principle consistent with the normalized strength of the observed bandheads. Especially the range in column densities spans orders of magnitude, while the fit results are similar within one order of magnitude. We return to this observation in the discussion.    

\section{Discussion} \label{sec:discussion}

\subsection{Comparison with previous CO bandhead modeling results} \label{sec:comparison}

We first compare the inner gas disk parameters determined from the CO bandhead emission with those resulting from similar modeling efforts in literature - particularly those that pertain to YSOs in the higher mass ranges. 

There are mainly two approaches when modeling CO bandhead emission towards YSOs. The first one assumes a thin disk model with free parameters similar to ours \citep[e.g.][]{kraus2000,wheelwright2010,ilee2013,ilee2014}. The second approach differs from the first mainly in that it does not fit a temperature or density structure, but models an isothermal ring at one density and $v\sin i$, where the velocity can be converted to a distance assuming an inclination and Keplerian rotation \citep[e.g.][]{bik2004,koutoulaki2019,fedriani2020,gravitycollaboration2020}. Both approaches have been used to successfully fit \first\,overtone emission. We do note that all previous studies normalize the bandheads to the peak of the first bandhead \citep[except][]{kraus2000,wheelwright2010}, thereby ignoring the strength of the emission. Furthermore, in our study, for the first time, the \second\,overtone CO bandhead emission is reported and fitted, leading to better constraints on the column density. With these differences in approach in mind, we summarize the similarities and differences in the results.

Our derived CO column densities are all of order $N_{\text{CO}}\sim 10^{21}$ \cmss, which is in excellent agreement with literature values that are mostly within the range $N_{\text{CO}}\sim 10^{20}-10^{22}$ \cmss. There is, however, often a larger spread in column densities when a sample of objects is studied \citep[e.g.][]{ilee2013,ilee2014}. While the morphology and strength of our bandhead profiles is equally diverse as in these studies, the determined densities within our sample are remarkably similar.

The spread in temperature values is similar to those found in previous studies. All inner radii in our sample are well within 1 AU, which is on the lower end of values derived for this parameter by others. \cite{ilee2013,ilee2014}, for instance, find inner radii up to 6 AU, but mostly around $1-2$ AU. They also generally find rather shallow temperature exponents ($p \sim -0.6$ on average), leading to (very) large outer radii (determined by where the temperature drops below 1000 K) ranging from a few to hundreds of AU. In our study, the strengths of the bandheads are very sensitive to $R_i$ and $p$ (\Cref{fig:comparison}), leading to (mostly) steep temperature exponents that mimic the effect of modeling a ring of emission close to the star. This justifies the assumptions underlying ring models. In the only study to date where the CO bandhead emission of a MYSO\footnote{NGC2024 IRS2} is spatially resolved, the authors indeed find, by fitting the visibilities for each bandhead separately, that the CO emitting region must be relatively small $\Delta R/R \leq 20\%$, where $R=0.58^{+0.04}_{-0.04}~\rm AU$ is the ring radius \citep{gravitycollaboration2020}. This radial location is in good agreement with our findings and the ring thickness is comparable to and even slightly narrower than our `rings' of emission (\Cref{tab:co_disk_props}). 

Similarly to other studies confronting the disk model to a sample of systems \citep{ilee2013,ilee2014}, the density exponent $q$ remains poorly constrained. This is not surprising since the emission region is narrow. This parameter is therefore also of little consequence to other results or statements about the disk (e.g. the total mass). 

For the inclinations we find a spread in values, consistent with random orientations. \cite{ilee2014} find a preference for higher (closer to edge-on) inclinations and suggest that this may be due to a geometric selection effect. This is interesting in the context of detection rates of CO bandhead emission (see \Cref{sec:kindofdisk}). Within our small sample, we do not find evidence for such a selection effect, in agreement with the larger sample of \cite{ilee2013}. In fact, we predict (see \Cref{fig:comparison}) that higher inclinations make the (optically thick) CO emission rather less detectable, due to a decrease in effective emitting surface. \\
\cite{carattiogaratti2017} and \cite{fedriani2020} report CO emission towards two deeply embedded MYSOs to be the reflection of light from the inner gaseous disk onto the perpendicular outflow cavity wall. In these cases the disk is almost fully edge on and the central source with its inner gas disk is too enshrouded for the CO emission to be detected directly. However, because the outflow cavity `sees' the inner disk face-on, the observed CO emission is rather fitted with low velocities (or a face-on orientation). It is highly unlikely that the CO emission towards our objects is indirect considering that all of them have detected photospheres. Thus, we do not expect that our inclinations are `artificially' low. But these observations can potentially help explain why detection rates of CO bandhead emission remain low overall: when looking at (embedded) MYSOs at high(er) inclinations one might not observe the innermost disk regions. 

Finally, the intrinsic line width $v_G$ (also denoted as $\Delta v$ in literature) is often taken to be a measure of turbulent motions of the gas. In our case, the main constraint on this parameter comes from line fluxes of the optically thick \first\,overtone, that become very high for higher widths ($>5~\rm km~s^{-1}$) that exceed thermal velocities of the gas at the given temperatures (\Cref{sec:model_exploration} and \Cref{fig:comparison}). Studies that fit the bandheads normalized to the first bandhead peak lack the sensitivity to this effect of intrinsic line widths. Such studies commonly find higher values up to $10-20~\rm km~s^{-1}$. 

In summary, our results point to very similar conditions for the origin of the CO emission as previous studies and, if anything, place even stronger constraints. 
Despite the fact that our objects represent different masses and (likely) different evolutionary stages (see \Cref{sec:kind_of_object}) than (M)YSOs investigated in the studies mentioned so far, it seems that the emission originates in disks that at least locally look very much alike. Even towards low-mass YSOs very similar conditions are derived, where the CO bandhead emission is known to signal a temperature inversion, i.e. a heated disk atmosphere by stellar irradiation, winds or magnetospheric accretion heating \citep[e.g.][see also \Cref{sec:kindofdisk}]{najita2007}. Thus, it is plausible to assume that in many if not all YSOs the emission originates in disk regions with similar conditions, if not (locally) similar disks.
\newcolumntype{C}{>{\centering\arraybackslash}m{0.056\linewidth}}
\newcolumntype{D}{>{\centering\arraybackslash}m{0.09\linewidth}}
\newcolumntype{O}{>{\centering\arraybackslash}m{0.165\linewidth}}
\newcolumntype{R}{>{\centering\arraybackslash}m{0.045\linewidth}}

\begin{table*}[ht!]
\footnotesize
\centering
\caption{Detection rates of CO \first\, overtone bandhead emission (CO emission) in YSOs of different masses and stages of formation. }  
\begin{minipage}{\hsize}
\centering
\renewcommand{\arraystretch}{1.4}
\begin{tabularx}{\linewidth}{D|O|C|C|X|C}
\toprule
Mass\footnote{The mass categories refer to expected ZAMS masses of approximately $\lesssim 2$ \msun (low), $2-8$ \msun (intermediate) and $\gtrsim 8$ \msun (high). Since mass estimates are not always available or derived in the same way the numbers are approximate.} & Object type & Detection rate & Sample size  & Comments & Reference \\
\midrule
Low & Class 0 & 67\% & 6 & & 1\\
Low & Class I & 15\% & 52 & & 2 \\
Low & Class I & 22\% & 110 & & 3 \\
Low  & T Tauri &  0\% & 100 & Sample from Lupus region. Authors find 3 cl TTS with CO emission in literature. & 4   \\
Low to intermediate & Variable, embedded (EXors, FUors, MNors) & 38\% & 28 &  60\% among objects identified as eruptive YSOs undergoing an episode of accretion. & 5  \\
Intermediate & Herbig AeBe & 5-21\% & 36 & & 6 \\
Intermediate & Herbig AeBe & 7\% & 91 & Strong preference for higher mass objects. & 7  \\
High & MYSO & 17\% & 195  & Red MSX Source (RMS) survey; low resolution spectroscopy. & 8 \\
High & MYSO (without \Hii\,region) & 34\% & 36  & Subsample RMS, selected to be luminous and radio quiet; intermediate resolution spectroscopy. & 9 \\
High & MYSO (with and without \Hii\,region) & 18\% & 11  & Sources associated with a disk (candidate) in sub-mm/radio and with a bright NIR counterpart. & 10 \\
\midrule
Intermediate to high & PMS stars in M17 & 63-83\% & 6-8 & Sample from which the objects in this study were taken.  & RT17 \\
\bottomrule
\end{tabularx}
\tablebib{(1) \cite{laos2021}; (2) \cite{doppmann2005}; (3) \cite{connelley2010}; (4) \cite{koutoulaki2019}; (5)\cite{contreraspena2017}; (6) \cite{ishii2001}; (7) \cite{ilee2014}; (8) \cite{cooper2013}; (9)\cite{pomohaci2017}; (10)\cite{hsieh2021}}
\end{minipage}
\label{tab:CO_detection_rates}
\normalsize
\end{table*}

\subsection{CO bandhead detection rates and the link with accretion and age}\label{sec:kindofdisk} 

CO \first\, overtone bandhead emission (in this section referred to as CO emission) has been observed in YSOs over a large mass range, as well as in Be and B[e] stars, which can be YSOs, but also objects of a more evolved nature such as B[e] supergiants \citep{lamers1998,kraus2020a} or classical Be stars \citep{cochetti2021}. Here we focus on findings for YSOs. 

We summarize the detection rates among YSOs of different masses and formation stages in \Cref{tab:CO_detection_rates}. Detection rates, especially among low mass YSOs, appear to be higher for earlier formation stages. In general, CO emission towards lower luminosity YSOs is associated with properties that point to the earliest stages of formation and/or high accretion rates, i.e. deep embedding, energetic outflows, veiling and/or measured high accretion rates \citep{najita2007}. In the intermediate to high mass YSOs detection rates are consistent with this trend. The lower rates among Herbig Ae/Be stars can perhaps be understood in terms of these objects representing a later stage in formation than most MYSOs, yet earlier than T Tauri's (see also \Cref{sec:kind_of_object}).

Apart from high occurrence rates in objects undergoing an accretion burst \citep[e.g.][]{contreraspena2017}, the link with accretion is supported by correlations of CO emission (strength) with other features that signal accretion, such as high veiling \citep{connelley2010} and Br-$\gamma$ luminosity \citep{ilee2014,pomohaci2017}. \cite{carattiogaratti2017} report on an accretion burst in a MYSO of $\sim20$ \Msun\, where the CO emission is detected \emph{only} during the burst. Despite all this evidence towards a link with accretion, actual reported accretion rates span a large range of values: from $\sim 10^{-8}~M_{\odot}\,\text{yr}^{-1}$ in the lowest mass objects \citep{koutoulaki2019} to $\sim 5\times 10^{-3}~M_{\odot}\,\text{yr}^{-1}$ in the highest mass objects \citep{carattiogaratti2017}, with Herbig Ae/Be stars in between with $\sim 10^{-7}-10^{-6}~M_{\odot}\,\text{yr}^{-1}$ \citep{ilee2014,contreraspena2017}. On the other hand, a theoretical study by \cite{ilee2018}  predicts CO emission to be most prominently observable with moderately high accretion rates of $\sim 10^{-4}-10^{-5}~M_{\odot}\,\text{yr}^{-1}$.

Finally, it should be noted that CO emission detection rates remain relatively low over all mass ranges, also in accreting objects. In MYSOs it is definitely less common than other emission features, such as Br-$\gamma$ emission at 2.16 \micron, also in the K-band, and which is detected towards 70-90 \% of objects in the quoted studies. Moreover, CO emission is not thought to be a direct probe of accretion, i.e. it does not originate in the accretion flow itself, where it is likely destroyed by UV light from the surface. As such it seems reasonable to follow \cite{koutoulaki2019} in their conclusion that accretion seems to be a necessary but not sufficient condition for CO emission to be observed. The low detection rates could be due to geometric effects (as discussed in \Cref{sec:comparison}), veiling due to high continuum emission or simply varying conditions in disks around accreting objects. It remains remarkable, however, that such vastly different accretion rates around objects of a wide range in masses could lead to emission that consistently points to very similar disk conditions (see \Cref{sec:comparison}). \newline

Our own sample was taken from a total of 6 YSOs, or 8, if we include the two young stars with only MIR excess from RT17 (B289 and B215 on \Cref{fig:hrd}; see also \Cref{sec:introduction}). It is surprising to find that 5 (63-83\%) of these objects show CO bandhead emission. This raises the question as to the evolutionary state of these objects, which we address in the following section.

\subsection{Evolutionary stage(s) of the intermediate to high mass YSOs in M17} \label{sec:kind_of_object}

The mass range of the YSOs in our sample  ($6-12$ \msun) spans the upper mass ranges of Herbig Be stars and the lower mass ranges of MYSOs. To better understand the formation stage of the studied sample, we briefly discuss the most important characteristics of the two categories and make comparisons to objects in each category.

Massive YSOs are minimally understood to be objects that are in the process of forming a star (or more in a multiple system) that is at least $8$ \msun\, when reaching the zero age main sequence (ZAMS). The widely used MYSO catalog Red MSX survey \citep[RMS;][]{lumsden2013} selects objects based on their MIR brightness and high luminosities ($\gtrsim 10^4~\text{L}_{\odot}$), which are hence defining properties for studies in the near- and mid-infrared \citep[e.g.][]{ilee2013,frost2019}. Some of these studies explicitly exclude from their definition objects that have started to ionize their surroundings to form an \Hii\,region \citep[e.g.][]{wheelwright2010,pomohaci2017}. However, this specification is not included in studies in (sub)-mm to radio wavelengths \citep[e.g.][]{beltran2016,maud2018,johnston2015,ilee2018a}, that reveal large scale ($\sim 10^2-10^4$ AU) molecular outflows and/or disk-like structures. \cite{takami2012} find that these mm-bright objects can be embedded at MIR wavelengths. Studies that explicitly include both wavelength regimes are rare \citep{hsieh2021}, which makes it hard to compare objects in either categories. It appears that the term MYSO is used for objects with a range of masses and evolutionary stages, from the near molecular core phase up to objects for which (some) photospheric features are observed, covering thereby a wealth of objects that differ in their characteristics. However, we note that the overwhelming majority of studies into MYSOs pertains to objects that are highly embedded in gas-dust envelopes ($A_V \gtrsim 20-30$) and have (estimated) masses of at least $10-15$ \msun\, with no detectable photospheric features. 

\begin{figure}
\hspace{-0.05\linewidth}
   \includegraphics[width=1.08\linewidth]{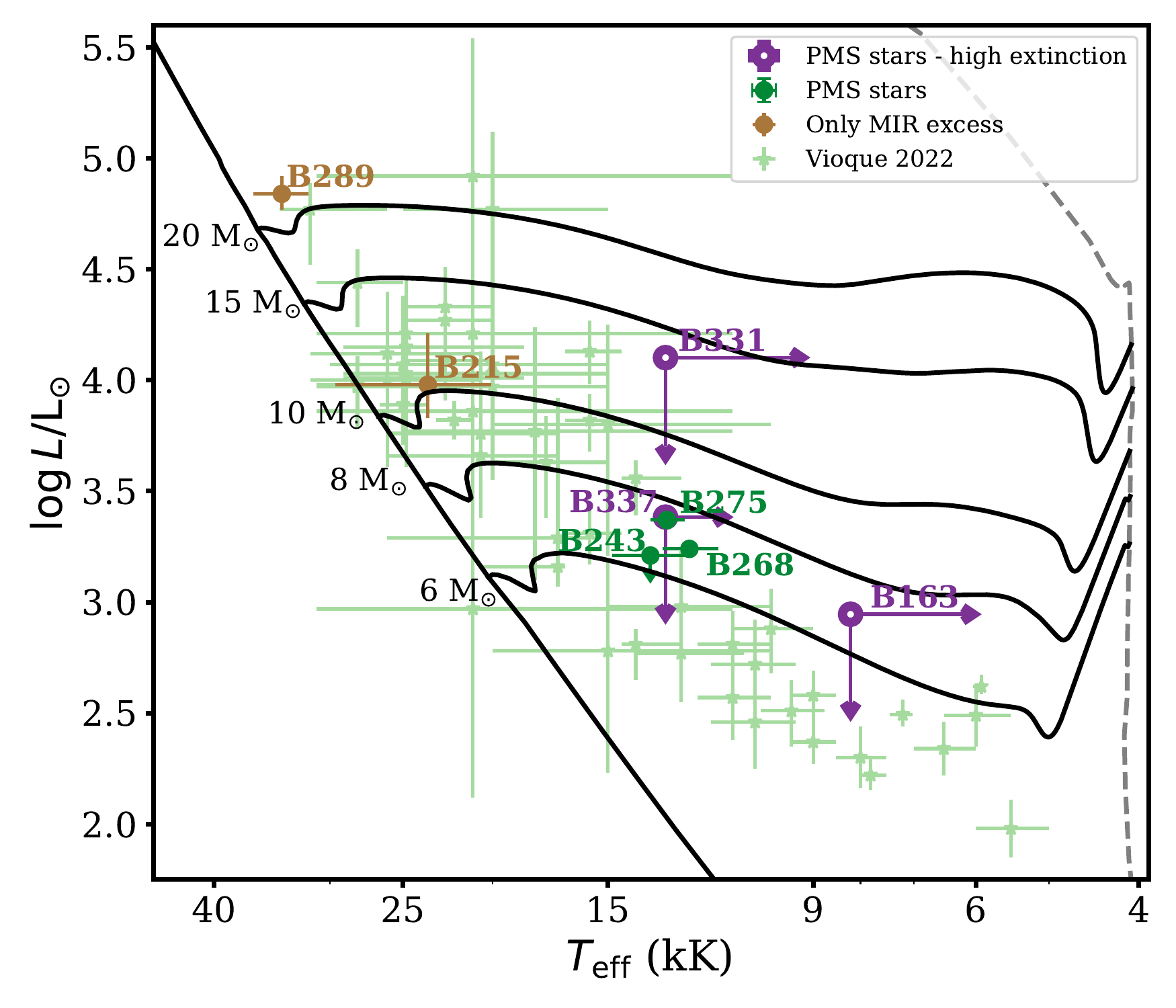}
      \caption{Hertzsprung-Russell diagram displaying the MIST pre-main-sequence tracks \citep{dotter2016} for stellar masses between 6 and 20 \msun, with the solid black line on the left denoting the ZAMS and the gray dashed line on the right indicating the birthline. The objects in this study, along with the one M17 PMS star from RT17 without CO bandhead emission (B337), are shown together with the $> 4$ \msun\, Herbig stars from the \cite{vioque2022} sample. Also included are the two RT17 objects (B289 and B215) close to the ZAMS without inner dust or gaseous disk signatures, but with MIR excess. The M17 PMS stars appear to represent a younger stage of formation, consistent with the high detection rate of CO bandhead emission in their spectra and their high extinction. For comparison, the estimated CO detection rate representative for $> 4$ \msun\, Herbig stars is $13-17\%$ (see text).}
         \label{fig:hrd}
\end{figure}

Of the M17 PMS stars, B331 with 12 \msun\, best fits the category MYSO, while also having a well detected photosphere. In a detailed study of a MYSO of $\sim 25$ \msun, \cite{frost2019} find evidence for a 60 AU inner clearing of the dust disk, while also reporting on the detection of CO bandhead emission. This combination of features resembles our findings for B331. Notably the authors refer to transition disks in low mass young stars which show dust clearing while still having a small gaseous disk where accretion could be ongoing \citep{wyatt2015}. They tentatively suggest that the object of their study could represent such a stage in an MYSO and propose photo-evaporation or a companion as the most likely dust clearing mechanisms. These conclusions are corroborated in their follow-up study of a sample of 8 MYSOs, from which they find evidence for an evolutionary sequence, the later stages of which are indeed characterized by inner dust clearing by photo-evaporation \citep{frost2021a,frost2021}. According to this interpretation the disk of B331 is in a state of transition. For our other objects this suggests an earlier stage of formation where they have not yet started to clear their inner dust disk. This scenario is consistent with the masses and luminosities of the respective objects. The two young stars from RT17 close to the ZAMS, with masses of 20 \msun\, and 10 \msun\, (B289 and B215 on \Cref{fig:hrd}), also support this picture as they lack all inner (gaseous or dust) disk signatures, but do show MIR excess similar to B331.  \newline

Recent studies that identify and characterize large samples of Herbig Ae/Be stars describe these objects as optically revealed PMS stars in the mass ranges $2-10$ \Msun, characterized by emission lines and NIR excess attributed to disks \citep[][]{vioque2018,vioque2022,guzman-diaz2021}. Because the M17 PMS stars fit this description well, we highlight the recent study of \cite{vioque2022} (hereafter V22)  for comparison. V22 analyze a sample of 128 Herbig Ae/Be stars with stellar masses up to $\sim 20$ \Msun. Based on optical spectra, GAIA parallaxes and photometry they derive stellar parameters and accretion rates. 

The V22 Herbig stars with masses $> 4$ \msun, most of which are spectral type B, are shown on the HRD in \Cref{fig:hrd} together with the M17 PMS stars. In the figure we include B337, the one PMS star studied by RT17 without CO bandhead emission, but with otherwise very similar characteristics. We also include the two objects from that study (B289 and B215; in brown), that show no signatures of inner (gaseous or dust) disks in their X-shooter spectra. The more extincted M17 sources ($A_V \sim 13$) are shown in purple. From the HRD plot, it appears that the M17 PMS objects are all younger than the V22 sources. This is consistent with the derived extinctions for the respective samples; with $A_V = 2.0-6.7$, and mean $\overline{A_V}=3.6\pm1.1$ for the V22 sub-sample ($M_* > 4$ \msun), and $A_V \sim 7 -13$ for the M17 sources (RT17 and \Cref{tab:stellar_properties}). Furthermore, V22 report photo-evaporation of inner dust disks for sources above 7 \Msun\, based on a decline in NIR/MIR excess, similar to B289 and B215.  In this context, the youth of the M17 PMS sources is again confirmed by the presence of inner dust and gaseous disks, with the most luminous and highest mass object (B331) showing signs of inner dust disk clearing \citep[in line with the comparison to the MYSO from][]{frost2019}. 

V22 do not mention CO bandhead emission, as the necessary wavelength ranges were not included in their study. The main reference for the occurrence rate of CO bandhead emission amongst Herbig Ae/Be stars therefore remains \cite{ilee2014}. Of the 23 objects in their sample with masses $\gtrsim 4$ \msun, $13-17\%$ exhibit CO bandhead emission \citep[masses from][]{fairlamb2015}. This is on the low side compared to detections towards MYSOs, which, as proposed earlier, could be because these objects are typically in a later stage of formation than most MYSOs. Though we did not perform a full census of PMS stars in M17, the presence of CO bandhead emission in 5 out of 6 them is yet another pointer towards young age.

\subsection{Limitations and open questions} \label{sec:limitations}

In this final section we discuss the limitations of this work and make suggestions for further research.

The column densities derived from the SED fitting (\Cref{tab:sed fit ranges}) are orders of magnitude lower than those derived from the CO bandhead fitting (\Cref{tab:co_best_fits_13CO}). The explanation for this lies in the fact that the disk model used lacks vertical disk structure. This means that the relatively high dust temperatures associated to the NIR emission are likely valid only on the disk surface, which can hide a cold and dense mid-plane underneath. Thus the derived column densities from the dust are representative only of the top disk layers and it is highly unlikely that the thermal emission is truly optically thin throughout the vertical disk extent as the best fitting models suggest. 

Similar statements can be made for the CO bandhead modeling: we have derived the properties of the region in the disk where the emission originates, which, as we have concluded, is relatively small. Therefore, it is difficult to make statements about the disk as a whole. Longer wavelength data (MIR and (sub)mm) are needed to constrain the masses and sizes of these disks. This, in turn could give further information on the disk evolution around these PMS stars - for instance whether the disks are (also) eroded outside-inwards. 

The question whether and how much these stars are accreting is also an open one. V22 estimate accretion rates of $\sim 10^{-4} - 10^{-6}$ \Msun/yr for objects in the mass range $6-12$ \Msun, based on line luminosities and extrapolations of magnetospheric accretion models. Since the M17 PMS stars appear younger than the V22 sample \emph{and} show CO bandhead emission, they could well be still accreting with rates on the higher end. However, since the similarity in disk properties from CO emission seems hard to reconcile with the disparity in reported accretion rates, the link with accretion remains ambiguous.

With regard to the possible accretion mechanisms, i.e. boundary layer (BL) vs. magnetospheric accretion, we note that although CO emission probes regions close to the star, for most of our objects it still originates from outside the corotation radius. With \vsini\,values taken from RT17 and using equation 2 from \cite{ilee2014} we find corotation radii of $\leq 0.081, 0.19 \text{ and } 0.14$ AU for B243, B268 and B275 respectively. Though we have no constraint on \vsini\,values for B163 and B331, the CO emission in those objects also (mostly) originates outside these radii. Thus, only for B268 (and perhaps B163) the CO emitting region covers the corotation radius. This could be an indication that the disk extends close enough to the stellar surface for BL accretion to take place, but for most of our objects the CO emission cannot probe this, in agreement with \cite{ilee2014}.

\section{Summary and conclusions} \label{sec:conclusions}
In this work we have studied the CO bandhead and thermal infrared emission from the disks around 5 intermediate to massive PMS stars in M17, with the aim to constrain their inner dust and gaseous disk properties and further our understanding of the end stages of massive star formation. To this end, we develop a LTE disk model that accounts for dust and CO overtone emission. For the first time we fit the $2^{\rm nd}$ overtone. We fit normalized bandheads, using models for stellar and dust continuum. We compare and discuss the derived disk properties in the context of the previously derived PMS evolutionary states of the central stars (RT17) and CO bandhead emission detection (rates) among YSOs in the literature. We arrive at the following conclusions:

\begin{enumerate}
    \item Taking into account the breadth of the parameter space and the diversity in line morphology, we arrive at surprisingly similar inner gaseous disk characteristics for the CO emitting region, which is consistent with results in the literature for YSOs of different masses. Therefore, it seems that this emission is typical of certain (more or less) specific physical conditions, i.e. high densities ($N_{\text{CO}}\sim 10^{21}~\text{cm}^{-2}$) and temperatures ($2000-5000$ K). 
    \item The (overall low) detection rates reported in the literature suggest that CO overtone bandhead emission is preferentially observed towards objects with high(er) accretion rates and in earlier stages of their formation, across all mass ranges. The high detection rate among the M17 PMS stars, in combination with their position on the HR-diagram suggests that these objects are exceptionally young for their mass range. 
    \item The previous conclusion is supported by the SED analysis, which points to undisrupted inner dust disks, except in the case of the most luminous object B331, for which the disk is likely in transition. 
    \item Though it is likely that these objects are accreting, the CO emission is not suited to probe the rate and mechanism of accretion. Results in this work point to CO bandhead emission regions close to the star, but outside the co-rotation radius (save for one object) within which boundary layer vs. magnetospheric accretion can be probed \citep[in agreement with][]{ilee2014}. Analysis of hydrogen lines for these objects is more suited to probe accretion regions, as these lines originate even closer to the stellar surface \citep{backs2023}.
    \item Taking into account the strength of the bandheads as well as the inclusion of the \second\, overtone emission provides more stringent constraints on the emission region and column densities, than previous studies lacking these diagnostics. 
    \item We clearly detect the \thirteenco~feature, but find no evidence for its enhancement in the spectra, consistent with the PMS nature of the M17 objects. Though of low diagnostic value to constrain disk parameters, we corroborate that \thirteenco\,abundances can be useful in determining the evolutionary state of objects with CO overtone emission, as suggested by \cite{kraus2020a}.
    \item The M17 intermediate to high mass YSOs appear unique in their combination of having both observable photospheres while also spectral and SED properties typical of younger, more enshrouded objects. Multi-wavelength analysis applied to the same objects is crucial for a better understanding of these objects in particular and the different stages of massive star formation in general. 
\end{enumerate}

\begin{acknowledgements}
We express our gratitude to the anonymous referee for their helpful comments and insights towards improving this manuscript. J. P. acknowledges support from the Dutch Research Council (NWO)-FAPESP grant for Advanced Instrumentation (P.I. L. Kaper). This work is based on observations collected at the European Organization for Astronomical Research in the Southern Hemisphere under ESO programs 089.C-0874(A) and 103.D-0099.
We thank SURF (www.surf.nl) for support in using the Lisa Compute Cluster. This research has made use of NASA's Astrophysics Data System Bibliographic Services. This work depended for a major part on the use of the Python programming language, in particular the packages NumPy \citep{harris2020}, SciPy \citep{virtanen2020}, and Matplotlib \citep{hunter2007}.
\end{acknowledgements}

\bibliography{co_disk_paper}

\begin{appendix}
\section{Photometry tables} \label{app:phot_tables}
\Cref{tab:denis_2mass,tab:wise_spitzer,tab:extra_phot} list the photometry points used for the fits. 
\begin{table}[h!]
\footnotesize
\centering
\caption{DENIS (I) and 2MASS (JHK) wavelengths and photometry (in magnitude.)}  
\begin{minipage}{\hsize}
\centering
\renewcommand{\arraystretch}{1.4}
\setlength{\tabcolsep}{3pt}
\begin{tabular}{lcccccccc}
\hline\hline
band 	& I       & $e_I$  &  J 	  &   $e_J$    &  H 	& $e_H$   &  K 	 &   $e_K$  \\
\hline
\diagbox[width=8em]{object}{wavelength}        & 0.791 \micron & & 1.235 \micron & & 1.662 \micron & & 2.159 \micron & \\
\hline
B163    & ...    & ...  &  12.877 &   0.052  &  10.915  & 0.036 &  9.686 &  0.027 \\
B243    & 14.907 & 1.00 &  12.252 &   0.026  &  10.815  & 0.029 &  9.544 &  0.026 \\
B268    & 14.332 & 1.00 &  11.794 &   0.043  &  10.538  & 0.041 &  9.494 &  0.027 \\
B275    & 12.804 & 0.03 &  10.467 &   0.033  &  9.128   & 0.029 &  7.947 &  0.024 \\
B331    & 15.754 & 0.06 &  11.355 &   0.024  &  9.822   & 0.033 &  8.946 &  0.036 \\
\hline
\end{tabular}
\end{minipage}
\label{tab:denis_2mass}
\normalsize
\end{table}

\begin{table}[hb!]
\footnotesize
\centering
\caption{WISE ($3.4$ and 4.6 \micron) and Spitzer GLIMSPE (3.6, 4.5, 5.8 and 8.0~\micron) wavelengths and photometry (in magnitude).}  
\begin{minipage}{\hsize}
\renewcommand{\arraystretch}{1.4}
\setlength{\tabcolsep}{3pt}
\begin{tabular}{lcccccccccccc}
\hline\hline
band	&  3.4~\micron  &  $e_{3.4}$ &  3.6~\micron &  $e_{3.6}$ &   4.5~\micron & $e_{4.5}$  &  4.6~\micron &  $e_{4.6}$ &  5.8~\micron &  $e_{5.8}$ &   8.0~\micron &  $e_{8.0}$  \\
\diagbox[width=8em]{object}{wavelength}        & 3.3526 \micron &              &  3.550 \micron & & 4.493 \micron & & 4.6028 \micron & & 5.731 \micron & & 7.872 \micron & \\
\hline
B163    & 7.955   & 0.032  & 7.919  & 0.069  &  7.225  & 0.047  & 7.000  & 0.033  & 6.64   & 0.042  &  6.082  & 0.11    \\
B243    & 8.148   & 0.042  & 7.829  & 0.046  &  7.216  & 0.057  & 7.235  & 0.032  & 6.676  & 0.034  &  6.227  & 0.118   \\
B268    & 7.9743  & 0.022  & 8.129  & 0.053  &  7.71   & 0.051  & ...    & ...    & 7.404  & 0.134  &  ...    & ...     \\
B275    & ...     & ...    & 6.662  & 0.062  &  5.88   & 0.033  & 5.8771 & 0.023  & 5.346  & 0.038  &  ...    & ...     \\
B331    & ...     & ...    & 7.718  & 0.192  &  7.017  & 0.087  & ...    & ...    & 6.026  & 0.135  &  ...    & ...     \\
\hline
\end{tabular}
\end{minipage}
\label{tab:wise_spitzer}
\normalsize
\end{table}

\begin{table}[hb!]
\footnotesize
\centering
\caption{Extra photometric fluxes (in Jy) from literature for B275 and B331.}  
\begin{minipage}{\hsize}
\centering
\renewcommand{\arraystretch}{1.4}
\setlength{\tabcolsep}{3pt}
\begin{tabular}{lcccc | c}
\hline\hline
 wavelength	(\micron) & 9.8 & 10.53 & 10.6   & 11.7 & Reference \\
\hline
B275    & ... & ... &  1.9 (0.31) & ... & 1 \\
B331    & 1.5 (0.2) & 1.8 (0.3) & ... & 2.1 (0.1) & 2\\
\hline
\end{tabular}
\tablebib{(1)\cite{nielbock2001}; (2)\cite{kassis2002}}
\end{minipage}
\label{tab:extra_phot}
\normalsize
\end{table}

\clearpage

\section{Continuum normalization and impact of stellar parameters}
\label{app:impact_stellar_params}

\begin{figure}[hb!]
\centering
   \includegraphics[width=2\hsize]{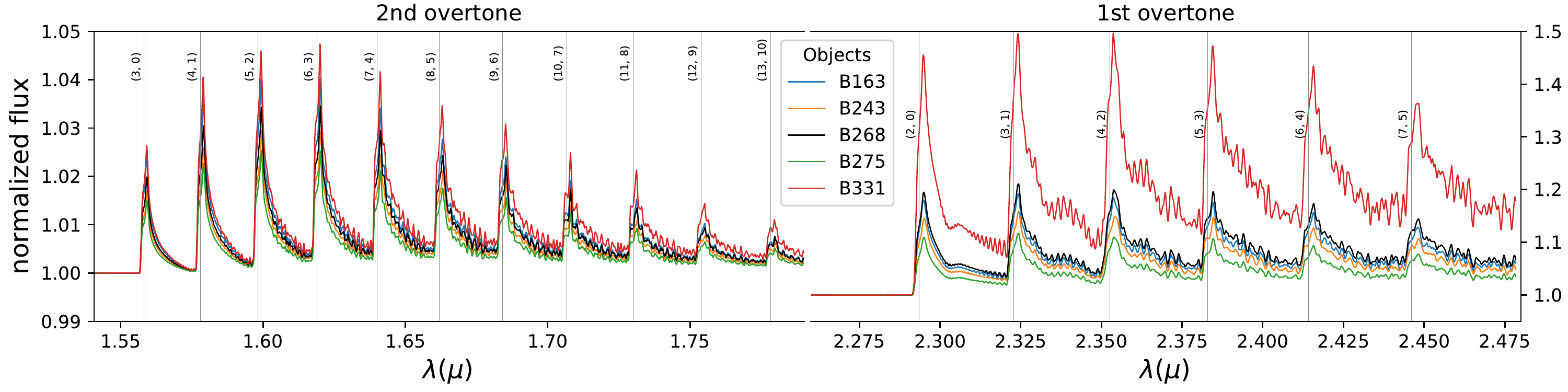}
      \caption[width=2\hsize]{Comparison of a model with the same disk parameters, changing only the object, i.e. the continuum, stellar mass and stellar radius. The model in black is a good fit to the B268 spectrum. Note the difference in scale for the $1^{\rm st}$ and $2^{\rm nd}$ overtone. The used disk parameters are: $T_i$ = 5000 K; $p$ = -0.75; $(N_\text{H})_i = 8.3\times 10^{25}~\text{cm}^{-2}$; $q$ = -1.5; $R_i$ = 1.1 $R_*$; $i$ = 50$^{\circ}$; $v_G$ = $2~\rm km~s^{-1}$; and \thirteenco~is not included.}
         \label{fig:comparison_stars}
\end{figure}

The effects stellar and continuum characteristics have on the appearance of the CO bandhead emission are visualized in \Cref{fig:comparison_stars}. Here we overplot the bandheads resulting from the exact same disk model for the 5 different objects in our sample. There are three main effects on the model fluxes:

\begin{enumerate}
    \item \textit{The continuum to which the bandheads are normalized.} 
    \newline This can been seen in \Cref{fig:comparison_stars} when comparing the $1^{\rm st}$ overtones of B243, B163, B268 and B275. The first three objects have similar SED fluxes in the relevant wavelength range (see \Cref{fig:sedfits}), while B275 has a much stronger continuum and therefore the weakest bandheads for the same model.
    
    In the same way, the relative strength of the continuum in the $1^{\rm st}$ and $2^{\rm nd}$ overtone wavelength regions also influences the relative strength of the respective set of bandheads. For B331 the slope of the SED is much steeper than for the other objects, causing the $2^{\rm nd}$ overtone for this object to be relatively weak.
    \item \textit{The stellar radius.} 
    \newline This is the measuring unit for $R_i$, which is the main reason that the $1^{\rm st}$ overtone of B331 is so much stronger: its stellar radius is simply much larger than that of the other objects (\Cref{tab:stellar_properties}), leading to a larger emitting surface for the same parameters. It also influences the Keplerian velocities which decrease as the square root of the stellar radius.
    \item \textit{The stellar mass.} 
    \newline Keplerian velocities are proportional to the square root of the mass. This effect, however, is largely compensated for by the previous point, as stars with higher mass generally also have larger radii. 
\end{enumerate}

\clearpage

\section{Fits and error calculations} \label{app:fits_errors}
\begin{figure*}[htb!]
\centering
   \includegraphics[width=1\hsize]{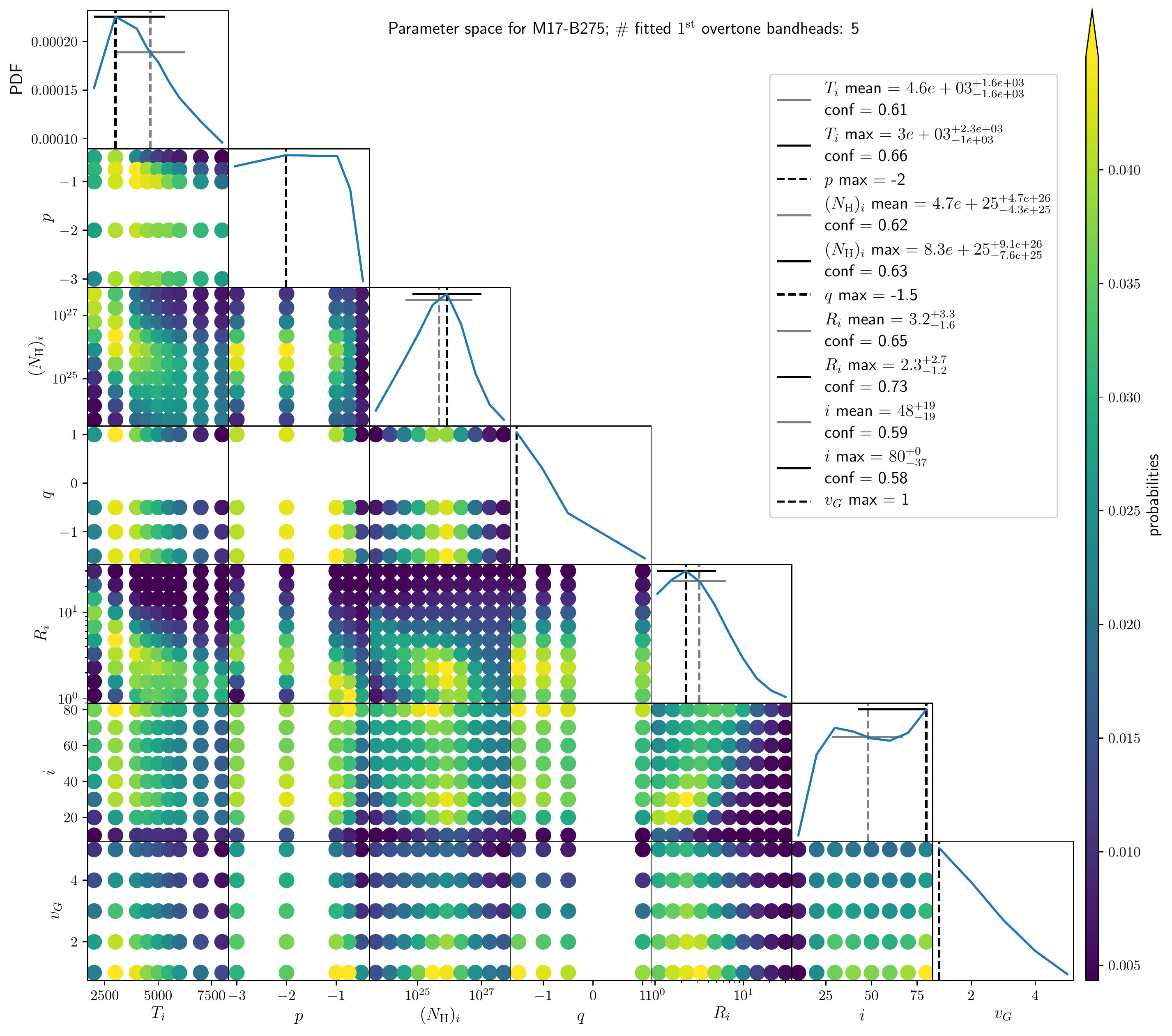}
      \caption{Off the diagonal: 2D marginalized likelihood distributions (see text), for the model grid fitting results on fitting five \first\, and six \second\, overtone bandheads for B275. On the diagonal: the 1D probability distribution functions (PDF) for each parameter resulting from normalizing the marginalized likelihood distributions to the area under the plotted line. The meaning of the mean, max and conf values is explained in the text.
      The patterns in the 2D distributions give insight into the degeneracies between parameters.} 
     \vspace{1pt}
         \label{fig:param_space_plot}
\end{figure*}

In order to get insight into the parameter space and to determine errors on the fit results, we made marginalized likelihood distributions for each parameter as follows. We converted all reduced $\chi^2$ values to a likelihood $L = e^{-\chi^2_{red}/2}$ and established the likelihood of a value $a$ for a parameter $p_i$ by summing over the likelihood of all other parameter value combinations:
\begin{equation}
    L(p_i=a) = \sum_{B_i}L(a;B_i)
\end{equation}
where $B_i$ is the set of all possible outcomes for all parameters except $p_i$. This likelihood was then normalized by the area under the curve (see \Cref{fig:param_space_plot} for visualization):
\begin{equation}
    P_{p_i}(a) = \frac{1}{\int_{-\infty}^{\infty} L_{p_i}(p_i')~dp_i'}  L_{p_i}(a)
\end{equation}
where the integration runs over the grid value range. The obtained 1D probability distributions $P_{p_i}$ have a mean:
\begin{equation}
    \mu = \int_{-\infty}^{\infty} p_i' P_{p_i}(p_i')~dp_i'
\end{equation}
and standard deviation:
\begin{equation}\label{eq:sigma}
    \sigma^2 = \int_{-\infty}^{\infty} (p_i'-\mu)^2 P_{p_i}(p_i')~dp_i' 
\end{equation}
where the integrations run over grid value range. Only if the resulting distribution is Gaussian, the best-fitting value and the maximum likelihood (or most probable) value are equal to the mean, which for most of our fits and parameters is not the case. An example of the resulting distributions is shown on the diagonal of \Cref{fig:param_space_plot}. The standard deviations on the most probable values (indicated as `max') are calculated as in \Cref{eq:sigma}, replacing $\mu$ with the maximum likelihood value. The confidence intervals indicated as `conf' are the 1-$\sigma$ confidence intervals obtained by integrating between $\mu - \sigma$ and $\mu + \sigma$. All errors were calculated such that no off-grid values fall within the 1-$\sigma$ confidence interval. Because of this and because the resulting distributions are not Gaussian, the 1-$\sigma$ confidence intervals does not generally result in (the Gaussian) $0.68$. Most intervals have values between $0.55$ and $0.75$.  On the off-diagonal panels in the figure the 2D equivalents of the described probability distributions are shown for each pair of parameters. The probabilities are color-coded with lighter colors indicating higher probability for a pair of values. These plots give insight into degeneracies in the parameter space and show how well a parameter is constrained.

The value that is quoted as the final result in \Cref{tab:co_best_fits_13CO} in this work, is the most probable or the mean value, depending which was closest to the best fit.

\end{appendix}
\end{document}